\documentclass[twocolumn,showpacs,preprintnumbers,amsmath,amssymb,superscriptaddress]{revtex4}

\usepackage{graphicx}
\usepackage{dcolumn}
\usepackage{bm}
\usepackage{longtable}
\begin{document}


\title{Deformed band structures in neutron-rich $^{152-158}$Pm isotopes}

\author{S.~Bhattacharyya}
\thanks{Corresponding author}
\email{sarmi@vecc.gov.in}
\affiliation{Variable Energy Cyclotron Centre, 1/AF Bidhannagar, Kolkata 700064, India.}
\affiliation{Homi Bhabha National Institute, Training School Complex, Anushaktinagar, Mumbai-400094, India.}
\author{E.~H.~Wang}
\affiliation{Department of Physics and Astronomy, Vanderbilt University, Nashville, Tennessee 37235, USA.}
\author{A.~Navin}
\affiliation{GANIL, CEA/DRF-CNRS/IN2P3, Boulevard Henri Becquerel, BP 55027, F-14076 Caen Cedex 5, France.}
\author{M.~Rejmund}
\affiliation{GANIL, CEA/DRF-CNRS/IN2P3, Boulevard Henri Becquerel, BP 55027, F-14076 Caen Cedex 5, France.}
\author{J.~H.~Hamilton}
\affiliation{Department of Physics and Astronomy, Vanderbilt University, Nashville, Tennessee 37235, USA.}
\author{A.~V.~Ramayya}
\affiliation{Department of Physics and Astronomy, Vanderbilt University, Nashville, Tennessee 37235, USA.}
\author{J.~K.~Hwang}
\affiliation{Department of Physics and Astronomy, Vanderbilt University, Nashville, Tennessee 37235, USA.}
\author{A.~Lemasson}
\affiliation{GANIL, CEA/DRF-CNRS/IN2P3, Boulevard Henri Becquerel, BP 55027, F-14076 Caen Cedex 5, France.}
\author{A.~V.~Afanasjev}
\affiliation{Department of Physics and Astronomy, Mississippi State University, Mississippi 39762, USA.}
\author{Soumik ~Bhattacharya}
\affiliation{Variable Energy Cyclotron Centre, 1/AF Bidhannagar, Kolkata 700064, India.}
\affiliation{Homi Bhabha National Institute, Training School Complex, Anushaktinagar, Mumbai-400094, India.}
\author{J.~Ranger }
\affiliation{Department of Physics and Astronomy, Vanderbilt University, Nashville, Tennessee 37235, USA.}
\author{M.~Caama\~{n}o}
\affiliation{USC, Universidad de Santiago de Compostela, E-15706 Santiago de Compostela, Spain.}
\author{E.~Cl\'ement}
\affiliation{GANIL, CEA/DRF-CNRS/IN2P3, Boulevard Henri Becquerel, BP 55027, F-14076 Caen Cedex 5, France.}
\author{O.~Delaune}
\affiliation{GANIL, CEA/DRF-CNRS/IN2P3, Boulevard Henri Becquerel, BP 55027, F-14076 Caen Cedex 5, France.}
\author{F.~Farget}
\affiliation{GANIL, CEA/DRF-CNRS/IN2P3, Boulevard Henri Becquerel, BP 55027, F-14076 Caen Cedex 5, France.}
\author{G.~de~France}
\affiliation{GANIL, CEA/DRF-CNRS/IN2P3, Boulevard Henri Becquerel, BP 55027, F-14076 Caen Cedex 5, France.}
\author{B.~Jacquot}
\affiliation{GANIL, CEA/DRF-CNRS/IN2P3, Boulevard Henri Becquerel, BP 55027, F-14076 Caen Cedex 5, France.}
\author{Y.~X.~Luo}
\affiliation{Department of Physics and Astronomy, Vanderbilt University, Nashville, Tennessee 37235, USA.}
\affiliation{Lawrence Berkeley National Laboratory, Berkeley, California 94720, USA.}
\author{Yu.~Ts.~Oganessian}
\affiliation{Joint Institute for Nuclear Research, RU-141980 Dubna, Russian Federation.}
\author{J.~O.~Rasmussen}
\affiliation{Lawrence Berkeley National Laboratory, Berkeley, California 94720, USA.}
\author{G.~M.~Ter-Akopian}
\affiliation{Joint Institute for Nuclear Research, RU-141980 Dubna, Russian Federation.}
\author{S.~J.~Zhu}
\affiliation{Department of Physics, Tsinghua University, Beijing 100084, People's Republic of China.}

\date{\today}
\begin{abstract}

High spin band structures of neutron-rich $^{152-158}$Pm isotopes 
have been obtained from the 
measurement of prompt $\gamma$-rays of isotopically identified fragments produced in 
fission of $^{238}$U+$^{9}$Be and detected using the VAMOS++ magnetic spectrometer
and EXOGAM segmented Clover array at GANIL and also from the high statistics 
$\gamma$-$\gamma$-$\gamma$ and $\gamma$-$\gamma$-$\gamma$-$\gamma$ data 
from the spontaneous fission of $^{252}$Cf using Gammasphere. 
The excited states in $^{157}$Pm and those above the isomers in even-A Pm isotopes 
$^{152,154,156,158}$Pm have been identified for the first time. 
The spectroscopic information on the rotational band structures in 
odd-A Pm isotopes 
has been extended considerably to higher spins and the possibility of
the presence of reflection asymmetric shapes is explored. 
The configuration assignments are based on the results of 
Cranked Relativistic Hartree-Bogoliubov calculations.
From the systematics of bands in odd-A
Pm isotopes and weak population of opposite parity bands, octupole  
deformed shapes in neutron rich Pm isotopes beyond $N=90$ seem unlikely
to be present.

\end{abstract}


\maketitle

\section{\bf Introduction}

\label{sec:level01}

The many-body correlations in atomic nuclei can lead to 
various shapes other than those which are spherically symmetric. 
These include  reflection symmetric prolate or oblate shapes having 
quadrupole deformation, triaxial shapes having both quadrupole deformation 
and axial asymmetry and reflection asymmetric ``pear shape'' with 
static octupole 
deformation. 
The nuclei of rare earth region with $N=88-90$ are well known to have 
transitional character: from nearly spherical or weakly deformed shapes for 
$N<88$ to well deformed shapes for $N>90$.
A deformed region beyond $N=90$ is known from the systematics of the
first excited  states of even-even nuclei around $Z=55-66$~\cite{Casten}. 
The nuclei in the $A=140-150$ mass region are 
expected to have octupole collectivity. This is because
their Fermi levels lie between the  $f_{7/2}$ and $i_{13/2}$ neutron 
orbitals and between the $d_{5/2}$ and $h_{11/2}$ proton orbitals, 
which differ by $\Delta j = \Delta l = 3$; this leads to an increase
of octupole correlations ~\cite{Ahmad, Butler}. 
The first experimental signature of octupole collectivity
in this region was reported for neutron rich Ba isotopes~\cite{Leander85, Phillips}
and later on more definitively established in neutron-rich odd and 
even A Ba and Ce nuclei~\cite{Zhu95,Zhu99,Chen06,Brewer} and 
other rare earth nuclei~\cite{Sheline}. 
Direct evidence of static octupole deformation in neutron rich nucleus 
$^{144}$Ba has been reported recently~\cite{Bucher}.

There are several experimental fingerprints of static octupole deformation 
in nuclei which are discussed in details in Refs.~\cite{Ahmad,Butler,Afana95}. 
One of them is the presence of specific features of rotational
       bands. Alternating parity bands with positive and negative
       parity states forming the sequence $I^+$, $(I+1)^-$, $(I+2)^+$ ...
       appear in even-even nuclei \cite{Butler}. Parity doublet bands 
       are formed in odd and odd-odd nuclei. In these bands, parity 
       doubling leads to the appearance of the pair of the states  
       with spin $I$ but opposite parities \cite{Butler}. Note that a
       parity doublet band can be represented as a pair of alternating 
       parity bands with simplexes $s=+i$ and $s=-i$ \cite{Butler}.
 The opposite parity states of given simplex in parity doublet 
bands are typically connected by strong E1 transitions, which is another 
possible experimental signature of reflection asymmetric shape. However, 
the electric dipole moment is built from delicate balance of proton and neutron 
contributions \cite{Butler1}.  As a result, low B(E1) rates do not 
necessarily exclude static octupole deformation and vice versa. 
For example, the large B(E1) strengths in $N=92$ isotones $^{155}$Eu 
and $^{157}$Tb could also be explained without considering static 
octupole deformation~\cite{Hartley}. The behavior of alternating-parity
and parity doublet bands with spin offers another clue on the 
presence of static octupole 
deformation since the rotation could stabilize the octupole deformation
\cite{Naz92,Butler1,Ana1993}. Therefore, to understand the 
evolution of nuclear shape and nature of deformation of nuclei in this 
region, it is important to have more experimental data on high spin 
states with increasing neutron number.

The observation of the bands with features typical for
parity doublet bands
in odd-A and odd-odd nuclei in this region has also lead to various 
theoretical efforts to interpret these results in terms of static octupole 
deformation~\cite{Leander84, Afana95}. Shell correction methods based on 
reflection asymmetric Woods-Saxon model were also used to explain the 
experimental observation of local quenching in E1 strength 
in some of the cases~\cite{Butler1}. 

The heavier isotopes of rare earth region, which are mostly neutron-rich, 
can be accessed by the fission process.
The spectroscopy of heavy fission fragments provides the opportunity
to investigate deformation effects as a function of neutron number for
a particular isotope chain.
Pm nuclei, with a wide isotopic range, are good candidates to explore the 
evolution of deformation with neutron number and possible role of octupole 
deformation in this region. 
It may be noted that there are no stable isotopes of Pm and that the isotopes 
beyond $^{151}$Pm are known only through radioactive decay 
studies~\cite{William, Shibata, Taniguchi, Karlewski, Greenwood, MShibata}, 
proton transfer measurements~\cite{Burke, Lee} and spontaneous fission of 
$^{252}$Cf source~\cite{Hwang}. This is mainly due 
to non-availability of suitable target-projectile combination to produce 
the nuclei in fusion evaporation reaction with large cross section. 
The high spin states of neutron rich nuclei in this mass region can be efficiently 
produced in fission. Isotopic identification of nuclei is critical to 
carry out spectroscopy of these exotic nuclei. 
Usually the technique of high fold $\gamma-\gamma$ coincidences and 
the cross coincidence relationships among the heavy and light fragment 
partners of the fission process 
are utilised to assign the $\gamma$ rays to a particular isotope. 
But in the case of extreme neutron rich isotopes, for which not a single $\gamma$
transition is known, it is very difficult to assign the $\gamma$ rays 
to the level scheme of a particular isotope on the basis of only
$\gamma-\gamma$ coincidence. Furthermore, the presence of low-lying long-lived isomers
in certain nuclei hinders the prompt coincidence between $\gamma$-rays below and above 
the isomer. In such cases, even if some of the transitions below the isomer are known,
the band structure above the isomer cannot be obtained by using 
high fold $\gamma$ coincidence techniques. Thus,
a direct isotopic (A,Z) identification of nuclei is essential for an 
unambiguous assignment of $\gamma$ rays. 
The high spin experimental data on neutron rich Pm isotopes was
very limited prior this study.
In fact, no in-beam high spin spectroscopic measurements are available beyond $^{151}$Pm.

The bands with features typical for
parity doublet bands
extending to high spins have been found in odd-A Pm isotopes
for $N=86$ $^{147}$Pm~\cite{147pm}, $N=88$ $^{149}$Pm~\cite{Jones}
and $N=90$ $^{151}$Pm~\cite{Vermeer, Urban}. 
The possibility of the presence of a reflection asymmetric shape at $N=92$
has also been reported from the observation of enhanced E1 transitions between 
a couple of low lying states of an 
band structure (which has typical features of parity doublet bands)
in $^{153}$Pm measured in $\beta$-decay~\cite{Taniguchi}.
Although the band structures observed in $^{151}$Pm and $^{153}$Pm are very similar, 
the origin of the ground state band in these two nuclei are quite different. 
In $^{151}$Pm the band head of the ground band is 5/2$^+$ based on 
 the [413]5/2$^+$ configuration, originating from  the $g_{7/2}$ orbital, 
whereas, the ground band in $^{153}$Pm with larger prolate deformation is based on 
the configuration [523]5/2$^-$, originating from the high-j $h_{11/2}$ orbital. 
The rotational band based on the $\pi$5/2[532] state has also been identified 
in $^{155}$Pm from the spontaneous fission of $^{252}$Cf source~\cite{Hwang}, 
whereas no excited states in $^{157}$Pm are reported so far. 
Recently, $\mu$s isomers are observed in $^{158, 159,161}$Pm, produced by 
in flight fission of $^{238}$U beam~\cite{Yoko}. Few lower spin members 
of the rotational band in $^{159, 161}$Pm have also been observed in this study 
of delayed $\gamma$ ray spectroscopy of $\mu$s isomers.
It would be interesting to 
extend the level structure of neutron-rich Pm isotopes to higher spins to 
understand the deformation effects as a function 
of N/Z and explore which kind of nuclear shapes (reflection symmetric 
quadrupole deformed or reflection asymmetric octupole deformed) exist at 
higher N/Z ratios.  

For the neutron-rich odd-odd Pm isotopes,
the information about the excited states is rather scarce, mainly 
because of the presence of long lived low lying isomers. 
Even the excitation energies of the long lived isomers are not well determined in most 
of the cases. 
The spectroscopic information on the odd-odd neutron-rich Pm isotopes above $N=90$
are reported from $\beta$-decay or decay of isomeric transition 
studies~\cite{William, Shibata, Karlewski, Greenwood, Dauria, MShibata}. 
Thus, the information about the structure of the states above the 
long lived isomers of these odd-odd Pm isotopes can only be obtained by directly populating 
the high spin states via in-beam prompt spectroscopy. 
In the presence of a long lived isomer, even if some of the $\gamma$-rays 
below the high spin isomer are known, the assignment of $\gamma$-rays above 
the isomer of a particular isotope is challenging, as $\gamma$-$\gamma$ 
coincidences across a long lived isomeric state are difficult.  
A few  low-spin states in $^{152}$Pm are known from $\beta$-decay 
of $^{152}$Nd~\cite{William, Shibata} and the existence of two high spin isomers
(7.5 min (4$^{\pm}$) and 18 min ($\ge$ 6$^+$)) have also been identified from these 
studies. The excited states of $^{154}$Pm have been identified from the 
$\beta$-decay of $^{154}$Nd~\cite{Karlewski, Greenwood} and  the existence of
two isomeric levels of 2.8 min and 1.8 min with probable spin assignment 
of J$\le$3 were reported in Ref.~\cite{Dauria} from the observed $\beta$-decay
and associated log-{\it ft} values corresponding to the $\beta$-decay of these 
isomeric states to $^{154}$Sm. For $^{156}$Pm, an isomeric state at 150.3~keV 
(with a tentative J$^{\pi}$ of 1$^-$) was identified and found to de-excite to 
the ground state by a M3 transition~\cite{MShibata}. In Ref.~\cite{Hwang}
the $\gamma$ rays from the high spin states of $^{156}$Pm were reported, 
but, as discussed 
later in this manuscript, those $\gamma$ rays actually belong to $^{157}$Pm. 
No spectroscopic information above these isomeric states in 
$^{152,154,156}$Pm are known prior to the present work. 
In the case of $^{158}$Pm, the evidence of the existence of a $\mu$s isomer
 and the isomeric transition decay from that state
has recently been reported~\cite{Yoko}. No other excited states of $^{158}$Pm 
are known prior to the present study. 

In the present work, the $\gamma$ rays are detected with EXOGAM~\cite{Sim00}  
segmented Clover detectors, in coincidence with the detected fission fragment 
at the focal plane of VAMOS++ spectrometer~\cite{Rej11} with isotopic (A,Z) identification.
 Thus, the assignment of $\gamma$ rays to a particular isotope is unambiguous. 
Once the identification of $\gamma$-ray is ensured, the details of the 
level scheme at higher spin can be obtained from the high-fold $\gamma$ 
coincidence data from $^{252}$Cf fission using Gammasphere. 
In the present work the prompt $\gamma$ rays of neutron-rich Pm isotopes
have been reported up to $N=97$ and the results are interpreted in terms
of rotational band structures. 
The in-beam populations of the prompt transitions of $^{153}$Pm 
is reported for the first time and 
the level scheme has been extended considerably to higher spins upto (29/2$^-$). 
The rotational band of $^{155}$Pm has also been extended to higher spins compared 
to previous work~\cite{Hwang}.  
The transitions of $^{157}$Pm are reported for the first time 
in this paper. Some of these transitions were assigned to $^{156}$Pm 
in Ref.~\cite{Hwang} from the measurement of spontaneous fission of $^{252}$Cf. 
For the odd-odd Pm isotopes, the first in-beam prompt spectroscopy 
measurements of $^{152-158}$Pm are reported from the present work. 
The band structures above the high spin isomers in these nuclei
have been identified for the first time. The results 
are discussed from the systematics of band properties. 
The Cranked Relativistic Hartree-Bogoliubov calculations are also 
performed to understand configuration assignments.

The paper is organized as follows. Section \ref{Exp-sect} describes
the details of experiment and the analysis of experimental data. 
Obtained results and the level schemes for odd-A and even-A Pm nuclei
are presented in Sec.\ \ref{sec:level02}. The discussion of the
observables of interest and the interpretation of 
physical situation are carried out in Section\ \ref{sec:level03}. Finally,
Sec.\ \ref{Concl}  summarizes the results of our work.

\section{\bf Experiment}
\label{Exp-sect}

The measurements were carried out by using two complementary methods, namely 
(i) by direct identification of the fission fragments (A,Z) at the focal plane
of the large acceptance magnetic spectrometer VAMOS++ and detection of 
corresponding in-beam $\gamma$ rays in coincidence and (ii) by 
detection of high fold $\gamma$ coincidence data from $^{252}$Cf spontaneous fission. 
The first method improves the selectivity and sensitivity of the measurements by 
unambiguous identification of the particular isotopes and the second one 
facilitates the study of high spin level structure using high-fold $\gamma$ coincidence 
techniques. 
The power of combining the two sets of data has been demonstrated earlier 
in case of study of neutron-rich Pr isotopes~\cite{Wang}.

\begin{figure}[ht]
\includegraphics[width=1.0\columnwidth]{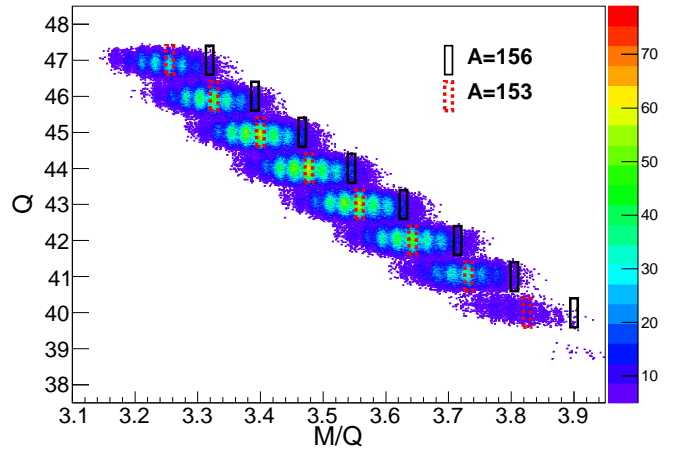}
\caption { \label{fig:id}  
(Colour online) 
Charge state (Q) as a function of the mass over charge (M/Q) after selection of  
Pm (Z=61) in the $^{238}$U+$^9$Be reaction at 6.2 MeV/u. The figure 
shows the identified Pm isotopes from various charge states at the focal 
plane of the VAMOS++ spectrometer. Identification of A=153 (red-dash box) 
and A=156 (black-solid box) from each charge 
state (Q) are labelled. The colour scale represents the counts along the Z-axis.} 
\end{figure}

The measurements of (A,Z) identification of the fission fragments and 
fragment-$\gamma$ coincidence were carried out at Grand Accelerateur National 
d'Ions Lourds (GANIL) using a $^{238}$U beam at 6.2~MeV/u ($\sim$0.2~pnA) on a 10-micron
thick $^{9}$Be target.  The fission fragments were directly identified by 
mass number (A), atomic number (Z) in the VAMOS++ magnetic spectrometer~\cite{Rej11} 
placed at $20^\circ$ with respect to the beam axis.
The elemental identification (Z) of the fission fragments were obtained from 
energy loss ($\Delta$E) in the isonization chamber and the total energy (E) measured 
by the Si detectors, placed at the focal plane of VAMOS++ spectrometer.
The Time-Of-Flight (TOF) was  measured between the two 
Multi-Wire Parallel Plate Avalanche Counters (MWPPAC), one placed just after 
the target and another at the focal plane. The (x,y) postions of the 
detected fragments were determined by the two Drift chambers at the focal plane.  
The various measured positions, energies and times  along with the
known magnetic field were used to determine, on an event-by-event basis,
the mass (M), charge state ($Q$), Z, and the velocity 
vector ($\vec{\rm{v}}$) for the detected fragment~\cite{Rej11}.
The magnetic rigidity (B$\rho$) was obtained by applying a reconstruction procedure. 
The parameters (M/Q) and (M) were obtained independently using the reconstructed
magnetic rigidity and from the measured velocity and total energy. 
A two-dimensional spectrum of Q an M/Q provides a clean and unambiguous 
identification of various isotopes detected at the focal plane. 
Fig.~\ref{fig:id} shows 
the identification plot of Q vs M/Q after selection of $Z=61$. 
The prompt  $\gamma$ rays were measured in coincidence with the isotopically-identified 
fragments, using the EXOGAM array~\cite{Sim00}, consisted of 11~Compton-suppressed 
segmented Clover HPGe detectors placed at 15~cm from the target position.
The  $\gamma$-ray energies of the fragments were obtained event by event 
after Doppler correction from the measured $\vec{\rm{v}}$ using the VAMOS++ spectrometer 
and with the known angle of the segment of the relevant clover detector~\cite{Rej11,Sam08}.
More details of the setup and the measurements at GANIL can be found 
in Refs.~\cite{Na14,Na14a}.

The high fold $\gamma$ coincidence data were obtained from the measurements of $^{252}$Cf
spontaneous fission at the Lawrence Berkeley National Laboratory (LBNL) using 101 HPGe 
detectors of Gammasphere. A 62-$\mu$Ci $^{252}$Cf source was sandwiched between two 
Fe foils of thickness 10 mg/cm$^2$. The data were sorted into $\gamma$-$\gamma$-$\gamma$
 and higher fold $\gamma$ events to form 3D cube and 4D cubes and were analyzed 
using the RADWARE package~\cite{Rad}. More details of this experiment and analysis
procedures can be found in~\cite{150Ce, 105Nb}.

\section{\bf Results}
\label{sec:level02}

\subsection{\bf Odd-A Pm isotopes}

\vskip 0.15cm
\begin{figure}[ht]
\includegraphics[width=\columnwidth]{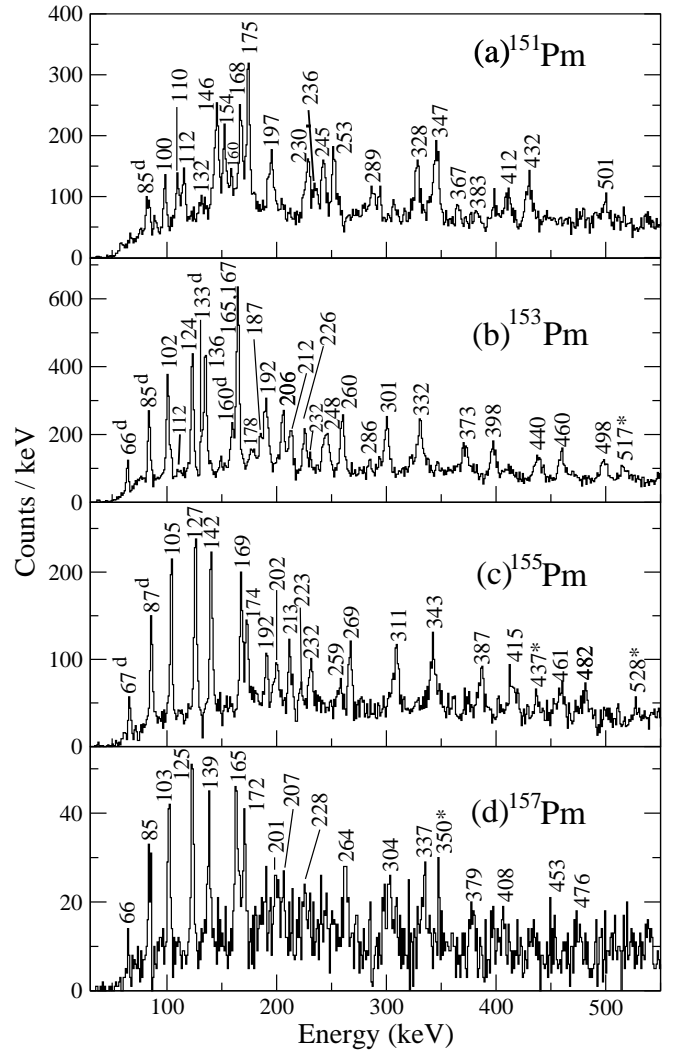}
\caption{\label{fig:spec-odd}(A,Z) gated Doppler corrected ''singles'' $\gamma$ spectra of 
odd-A $^{151-157}$Pm, obtained from the fragment-$\gamma$ coincidences in 
$^{238}$U+$^{9}$Be-induced fission data. 
The $\gamma$ rays earlier observed from $\beta$ decay are marked as 'd'.
The $\gamma$ rays marked as '*' are assigned to the respective odd-A Pm isotopes
from the corresponding (A,Z) coincidence, but could not be 
placed in the level scheme.} 
\end{figure}

\begin{figure}[ht]
\includegraphics[width=\columnwidth]{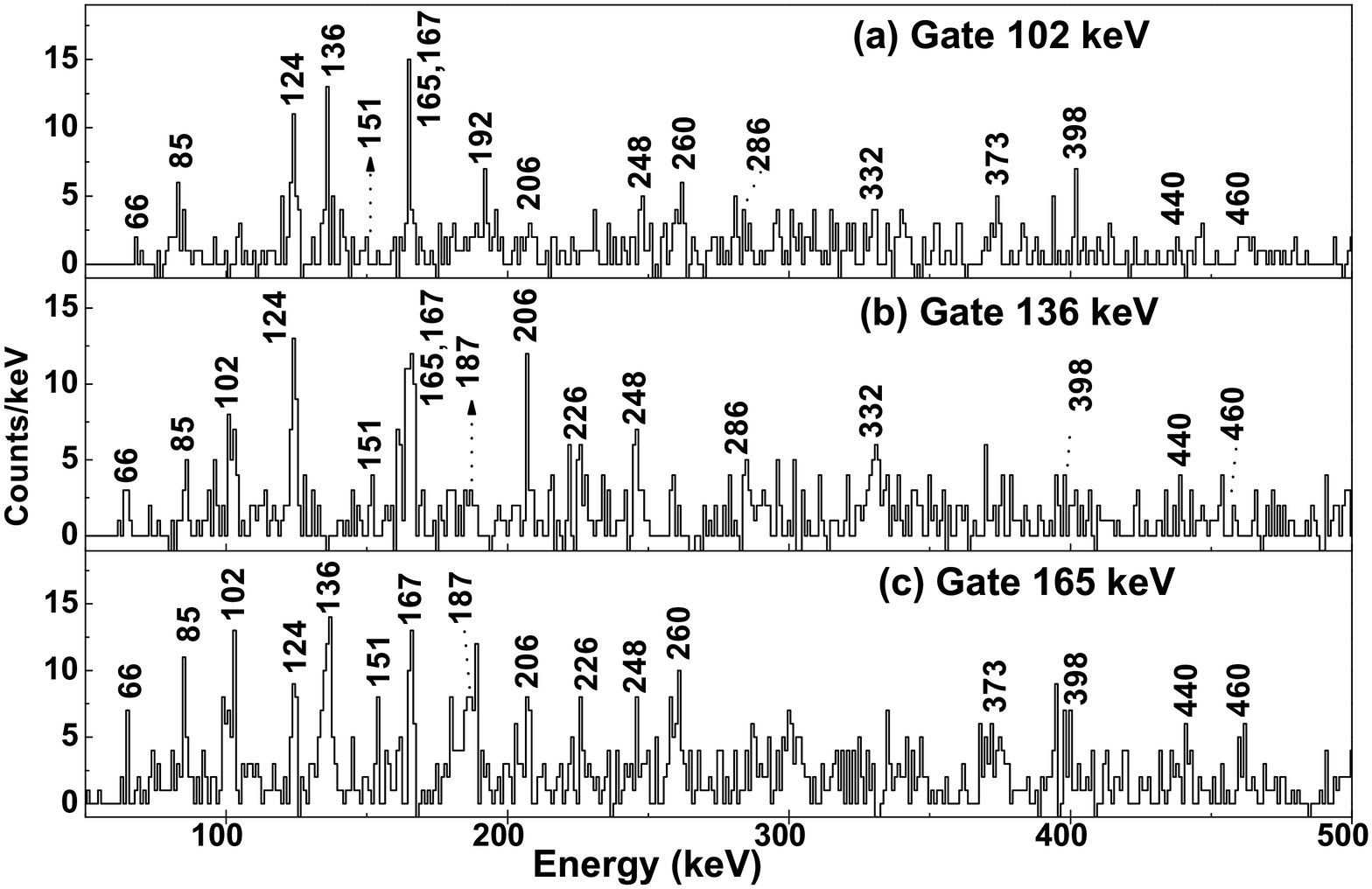}
\caption{\label{fig:gg-153}  
Coincidence spectra corresponding to the gates on the (a) 102, (b) 136 and (c) 165 keV 
transitions in $^{153}$Pm, obtained
from the $^{238}$U+$^{9}$Be-induced fission data. }
\end{figure}

\begin{figure}[ht]
\includegraphics[width=\columnwidth]{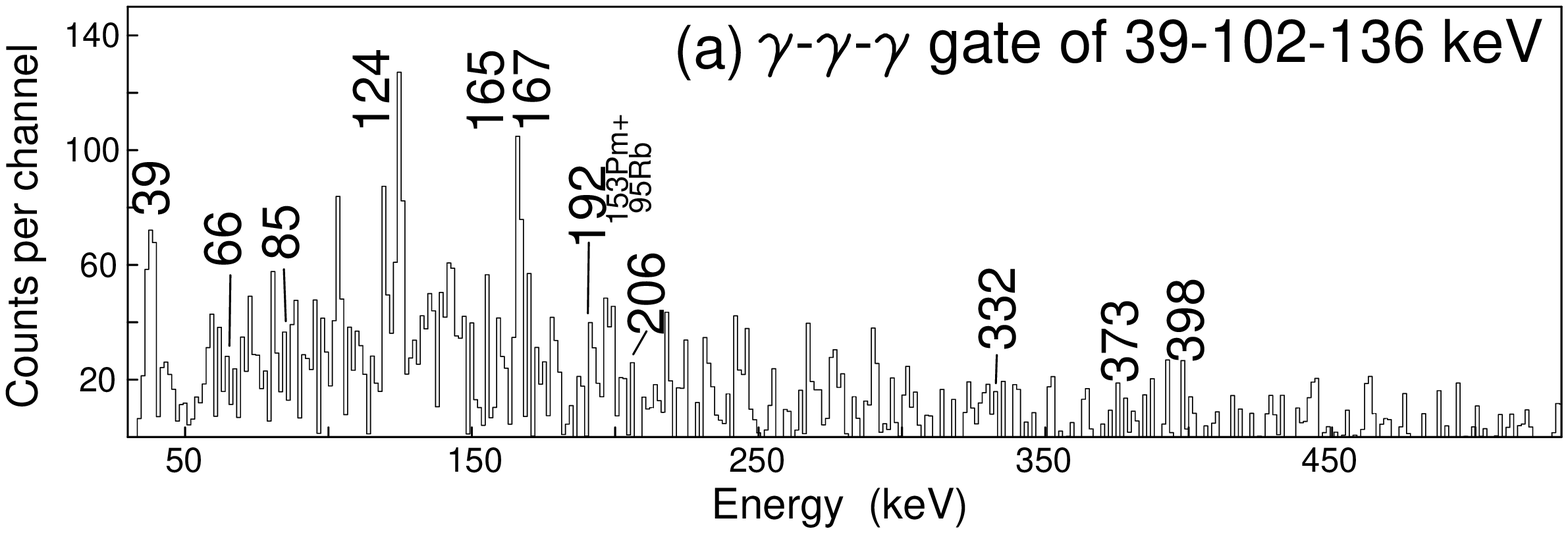}
\includegraphics[width=\columnwidth]{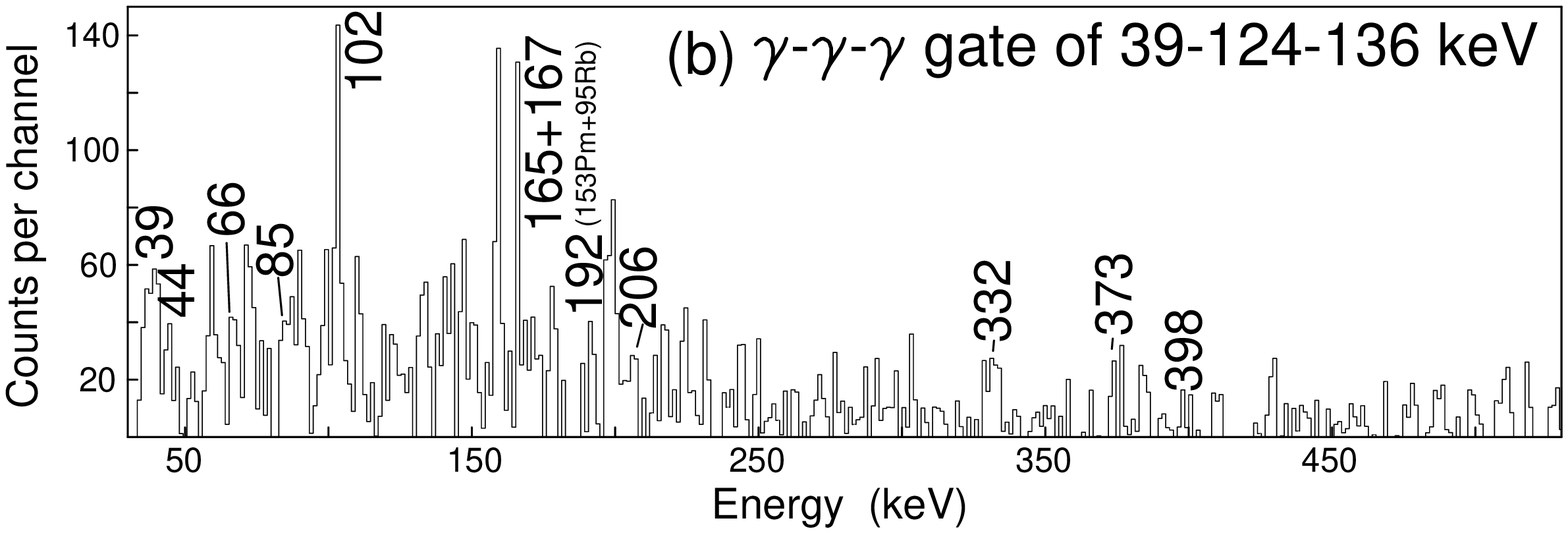}
\caption{\label{fig:tripple-153}  
 Coincidence spectra corresponding to triple gates of the 
transitions in $^{153}$Pm, obtained from the $^{252}$Cf spontaneous fission data. 
(a) 39 (Pm X-ray)-102-136 keV coincidence spectrum and 
(b) 39 (Pm X-ray)-124-136 keV coincidence spectrum. }
\end{figure}

\begin{figure}[ht]
\includegraphics[height=0.9\columnwidth, angle=-90]{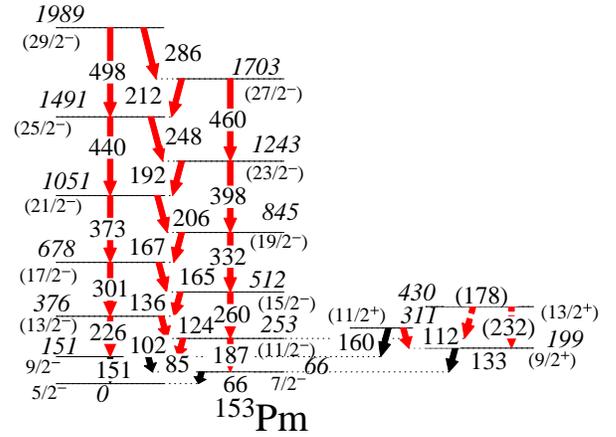}
\caption{\label{fig:scheme-153}  
(Colour online) The level scheme of $^{153}$Pm proposed in the present work.
The new transitions placed from the present work are shown in red colour and 
the previously known transitions from $\beta$-decay are shown in black.
The spin-parity of the levels are tentatively assigned  
from the systematics of odd-A Pm isotopes, except for 7/2$^-$ and 9/2$^-$ 
levels, which are adopted from Ref.~\cite{Taniguchi}. } 
\end{figure}

\begin{figure}[ht]
\includegraphics[width=\columnwidth]{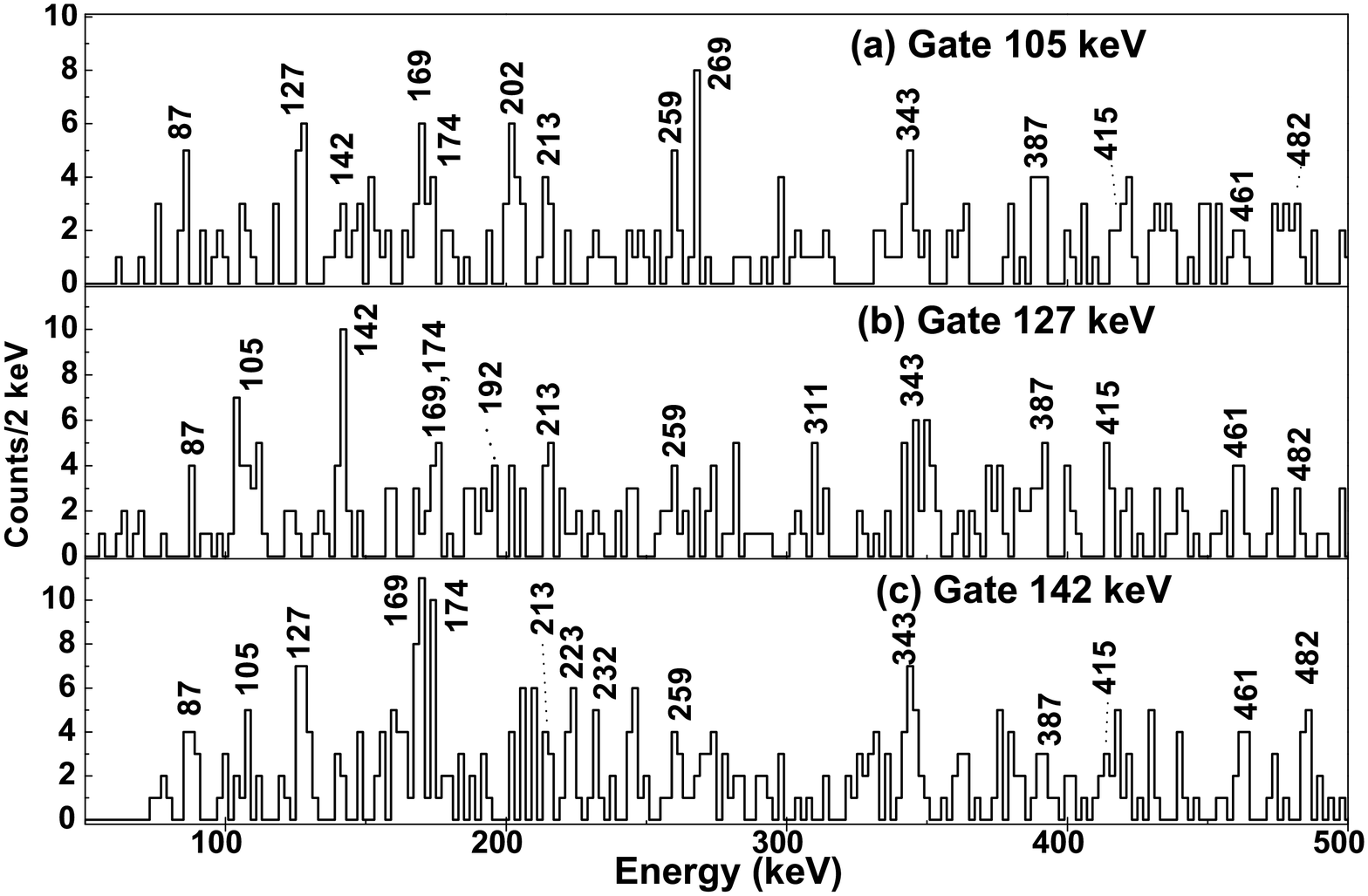}
\caption{\label{fig:gg-155}  
Coincidence spectra corresponding to the gates on the (a) 105, (b) 127 and (c) 142 keV 
transitions in $^{155}$Pm, obtained
from the $^{238}$U+$^{9}$Be-induced fission data.} 
\end{figure}

\begin{figure}[ht]
\includegraphics[height=\columnwidth, angle=90]{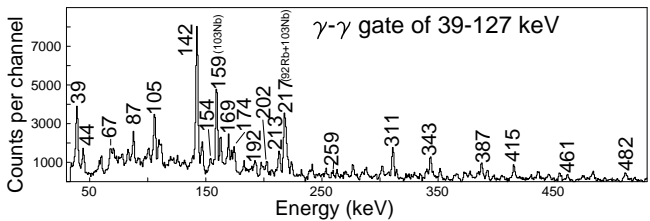}
\caption{\label{fig:tripple-155}  
Coincidence spectrum corresponding to the double gates of 39 (Pm X-ray)-127 keV  
transitions in $^{155}$Pm, obtained
from the $^{252}$Cf spontaneous fission data. }
\end{figure}

\begin{figure}[ht]
\includegraphics[height=0.7\columnwidth, angle=-90]{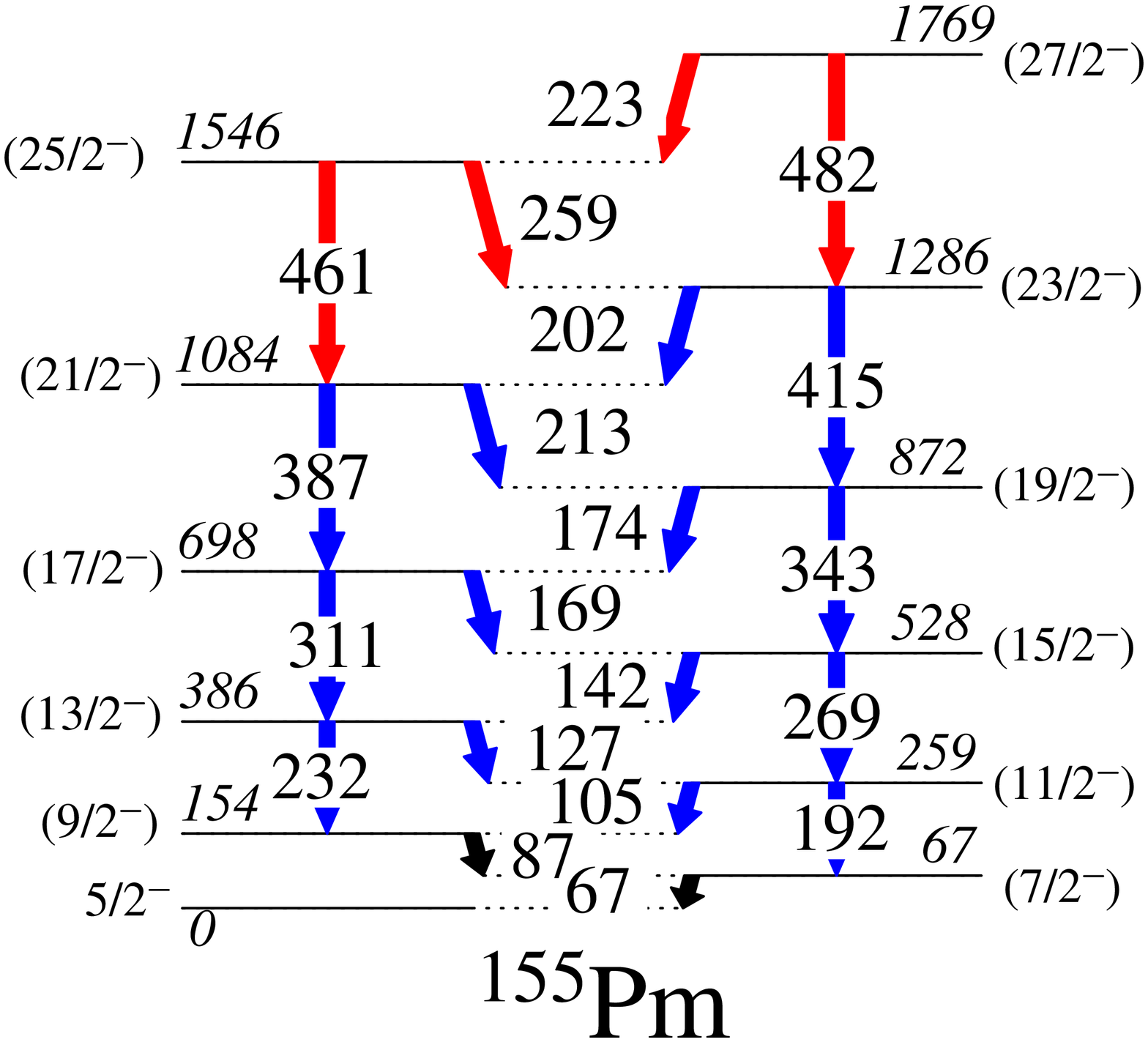}
\caption{\label{fig:scheme-155}  
(Colour online) Level scheme of $^{155}$Pm proposed in the present work. 
The new transitions obtained in the present work are shown in red colour.  
The transitions previously known from $\beta$-decay and $^{252}$Cf fission 
are shown in black and blue colours respectively.
The spin-parity of the levels shown are tentatively assigned  
from the systematics of odd-A Pm isotopes. }    
\end{figure}

Doppler corrected $\gamma$-ray spectra of odd-A Pm isotopes, 
detected in the present work by EXOGAM segmented Clover detectors, 
after the (A,Z) selection of the respective isotopes at the focal plane of 
VAMOS++ spectrometer, are shown in Fig.~\ref{fig:spec-odd}. 
All the transitions of $^{151}$Pm up to spin 27/2 $\hbar$, observed 
in the present work confirm those previously reported~\cite{Urban, Vermeer}. 
The corresponding spectrum is shown in Fig.~\ref{fig:spec-odd}(a) to highlight
the quality of data. 
For neutron rich odd-A Pm isotopes, transitions known from previous $\beta$-decay 
studies, are labelled. 
Some of the $\gamma$-rays (as shown in Fig.~\ref{fig:spec-odd}), 
though identified as belonging to the 
corresponding odd-A Pm isotopes through (A,Z) gating, could not 
be placed in the respective level schemes due to insufficient $\gamma$-$\gamma$ 
coincidence information.
The $\gamma$ rays of $^{153, 157}$Pm from in-beam prompt spectroscopy 
are being reported for the first time here 
and new level schemes are proposed. New transitions in  $^{155}$Pm have also 
been identified and the level scheme has been extended to higher spins compared 
to that previously reported~\cite{Hwang}. 
The $\gamma$ rays, assigned to various odd-A Pm isotopes, along with their 
relative intensities 
and the level energies, obtained from the (A,Z) gated spectra are shown in 
Table.~\ref{tab:Table1}. The quoted errors are the fitting errors. 
The probable spin-parity assignments of 
the initial and final states of the transitions have been obtained using  
the known systematics of odd-A Pm isotopes. The statistics was not enough 
to carry out any angular distribution or polarization measuremets 
for such neutron rich nuclei.

Information on the excited states of $^{153}$Pm was known from (t,$\alpha$) 
and (d, $^3$He) transfer reactions ~\cite{Burke, Lee}. 
The levels pertaining to the bands based on Nilsson orbitals 
5/2$^-$[532], 5/2$^+$[413], 3/2$^+$[411] were identified from the measured 
particle angular distributions, 
without detection of any decaying $\gamma$ transitions between the states.
First indication of 
band structure 
in $^{153}$Pm, which has typical features of parity doublet bands,
was reported from $\beta$-decay measurements~\cite{Taniguchi}.
In the present work the first information on the high spin states in $^{153}$Pm 
is obtained, which is extended up to 1989 keV (29/2$^-$) from in-beam prompt $\gamma$ 
spectroscopy measurements. 
The $\gamma$ spectrum in coincidence with the $^{153}$Pm fragments identified 
using the VAMOS++ spectrometer is  shown in Fig.~\ref{fig:spec-odd} (b). 
The $\gamma$ spectra corresponding to the gates of 102, 136 and 165~keV transitions
are shown in Fig.~\ref{fig:gg-153}, obtained from the $\gamma$-$\gamma$ matrix 
after selection of $^{153}$Pm fragment from $^{238}$U+$^{9}$Be-induced fission data. 
The mutual coincidence of 102, 124, 136, 165, 167~keV $\gamma$ rays can be 
clearly seen from the coincidence spectra shown in Fig.~\ref{fig:gg-153}(a-c). 
The presence of doublet 165-167~keV could be confirmed from the 
$\gamma$-$\gamma$ coincidence, as it is evident from Fig.~\ref{fig:gg-153}(c) 
corresponding to the coincidence spectrum of 165~keV transition. 
The presence of other transitions decaying from higher 
excited states of $^{153}$Pm, is also visible from the coincidence spectra of 
Fig.~\ref{fig:gg-153}(a-c). 
The triple coincidence spectra by gating on the 39~keV Pm-X-rays and 
other strong transitions 
of 102, 124 and 136~keV are shown in Fig.~\ref{fig:tripple-153}. 
The presence of cascade $\gamma$ rays in the 5/2$^-$ ground band of $^{153}$Pm
is clearly visible from these spectra.  
The level scheme of $^{153}$Pm, as shown in Fig.~\ref{fig:scheme-153}, has been obtained 
on the basis of energy systematics of known 
odd-A Pm isotopes, the $\gamma$-$\gamma$ coincidence measurements and intensity balance.
The placement of the crossover transitions are also made from the energy-sum systematics.
The spin-parity of the 7/2$^-$, 9/2$^-$ states are adopted from a 
previous work on $\beta$-decay~\cite{Taniguchi}. 
In the present work, the data statistics for each detector angle 
was insufficient to carry out an angular distribution analysis. 
Therefore, the spin-parities of 
the other excited levels are only tentatively assigned from the systematics 
of odd-A Pm isotopes. 
The observed intensities of transitions placed in the level scheme of $^{153}$Pm and the 
corresponding excited levels with tentative spin-parity assignments are given in  
Table.~\ref{tab:Table1}. It may be noted that, the low energy dipole transitions 
can be significantly converted and contain a mixing of higher order multipoles. 
As mixing ratios (from angular distributions) of the transitions or the 
conversion electrons could not be measured in the present experiment, 
therefore, only the observed $\gamma$ ray 
intensities are shown in Table.~\ref{tab:Table1}. 
The intensity mismatch for some of the lower
lying states (7/2$^-$, 9/2$^-$ and 11/2$^-$) can be due to the fact that the low energy 
dipole transitions are not pure and depending on the mixing ratios 
the conversion coefficients will be altered. 

The presence of a parity doublet band in $^{153}$Pm was indicated earlier in 
Ref.~\cite{Taniguchi}. The transitions pertaining to this band, as 
reported in Ref.~\cite{Taniguchi}, could also be confirmed from the present data. 
From the present measurement, 
the 112~keV $\gamma$ ray has been placed in this band as the transition from 
(11/2$^+$) to (9/2$^+$) level. 
Also, the 178~keV and 232~keV $\gamma$ rays have been tentatively placed 
in this band, as these two transitions could not be observed 
in coincidence with any of the 
strong transitions of the band based on (5/2$^-$) state. 
Though $\gamma$-$\gamma$ coincidence information could not be 
obtained for these transitions due to limited statistics, their presence 
in $^{153}$Pm is confirmed, as the $\gamma$-spectra in the present work 
have been obtained after (A,Z) identification. 

The spectroscopic information on the excited states of $^{155}$Pm and the 
associated $\gamma$ rays were first known from the study of the $\beta$-decay of 
$^{155}$Nd~\cite{Greenwood1}.
Later, the existence of a rotational band extending up to (23/2$^-$) in $^{155}$Pm 
was reported ~\cite{Hwang} from the measurements of $^{252}$Cf fission. 
This band has been extended up to (27/2$^-$) in the present measurements. 
The $\gamma$-ray singles spectrum, as obtained using EXOGAM Clover array, in 
coincidence with detected $^{155}$Pm at the focal plane of VAMOS++ spectrometer 
is shown in Fig.~\ref{fig:spec-odd}(c). The new transitions as compared to earlier 
reported~\cite{Hwang} are observed from this spectrum. 
The $\gamma$ spectra corresponding to the gates of 105, 127 and 142~keV transitions
are shown in Fig.~\ref{fig:gg-155}, obtained from the $\gamma$-$\gamma$ matrix 
after selection of $^{155}$Pm fragment from $^{238}$U+$^{9}$Be-induced fission data. 
The placement of 105, 127 and 142~keV $\gamma$ rays in coincidence with each other 
and the higher lying transitions are supported   
from the coincidence spectra shown in Fig.~\ref{fig:gg-155} (a~-~c). 
Coincidence of the transitions are checked using various gating conditions 
and also from different added gates. 
A coincidence spectrum 
corresponding to double gates of 39~keV (Pm X-ray) and 127~keV, obtained from 
$^{252}$Cf data, is shown in Fig.~\ref{fig:tripple-155}. In this coincidence 
spectrum, the 159 and 217~keV peaks appear due to the presence of 126-159-218-217 
cascade in $^{103}$Nb and its partner La K$_{\beta}$ X-ray at 38 keV, 
which is present as a contamination in $^{155}$Pm. 
The 217 peak could also come from $^{94}$Rb fission partner in this case. 
The energy sum, intensities and the coincidence information of the observed 
$\gamma$ rays are used to obtain the level scheme, as shown in  
Fig.~\ref{fig:scheme-155}. The spin-parities indicated in Fig.~\ref{fig:scheme-155} 
are only tentative, as angular distribution and 
polarization measurements were not possible from the present data. 
Also, it may be noted that no indication of any side band could be found 
for $^{155}$Pm from the present data. 

\begin{figure}[h]

\includegraphics[height=\columnwidth, angle=90]{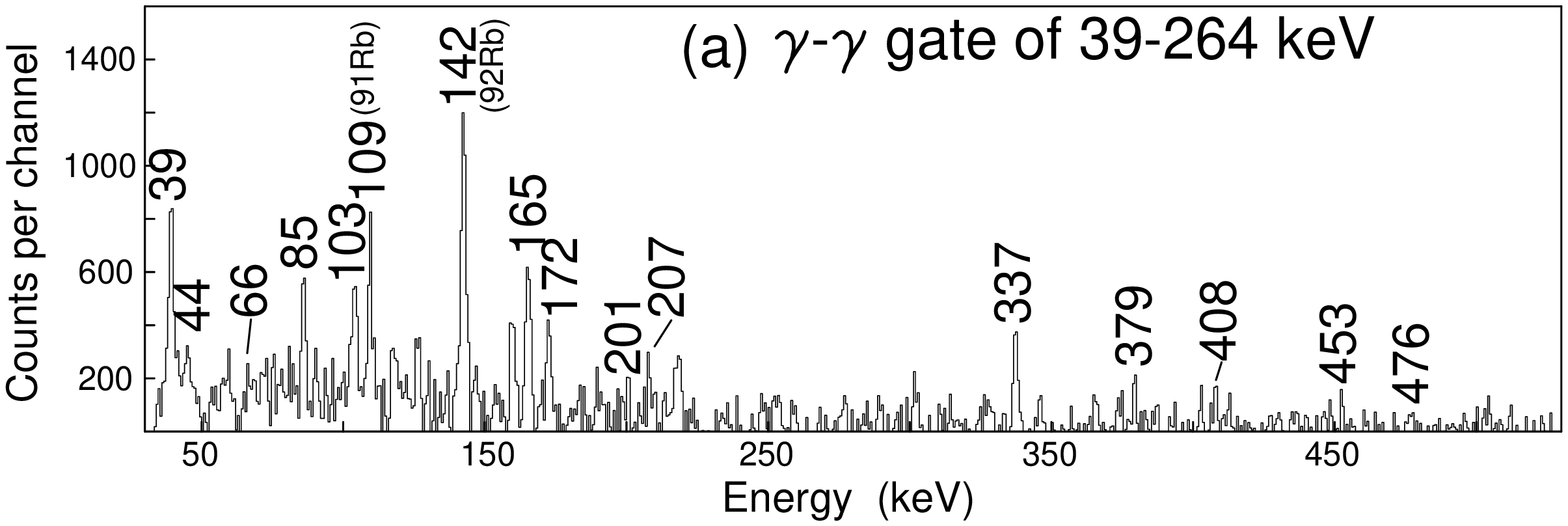}
\includegraphics[height=\columnwidth, angle=90]{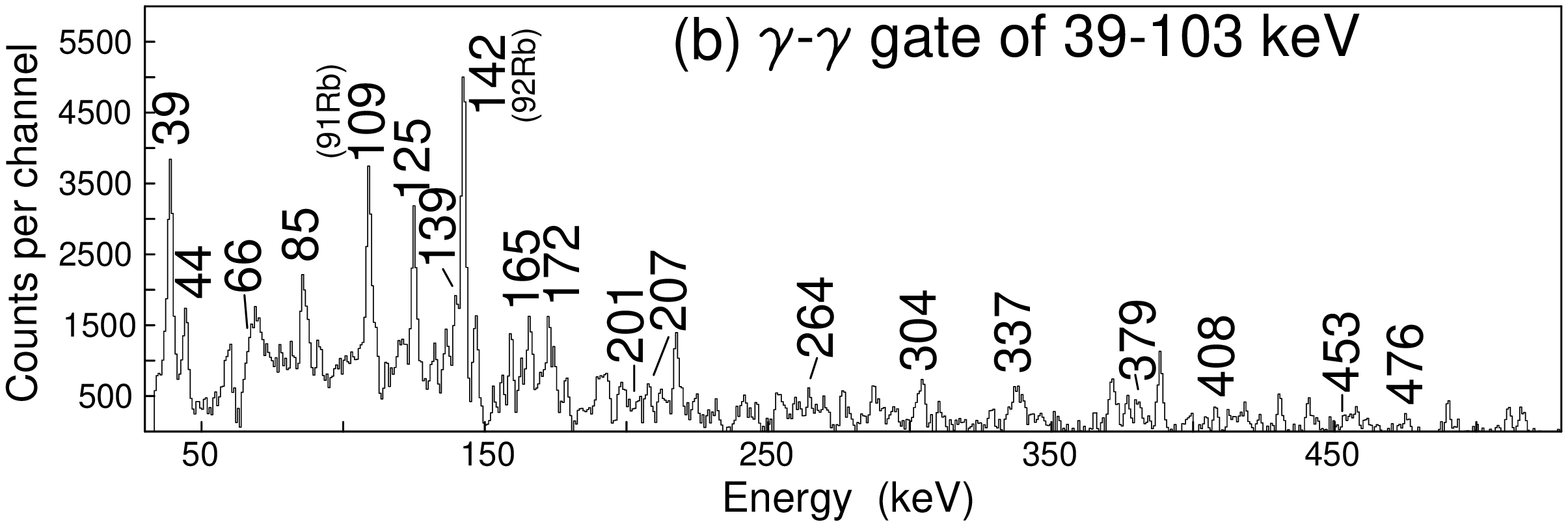}
\caption{\label{fig:tripple-157}  
Coincidence spectra corresponding to the double gates of the transitions in $^{157}$Pm, 
obtained from the $^{252}$Cf spontaneous fission data. 
(a) 39 (Pm X-ray)-264 keV coincidence spectrum and 
(b) 39 (Pm X-ray)-103 keV coincidence spectrum. }
\end{figure}

\begin{figure}[h]
\includegraphics[height=0.7\columnwidth, angle=-90]{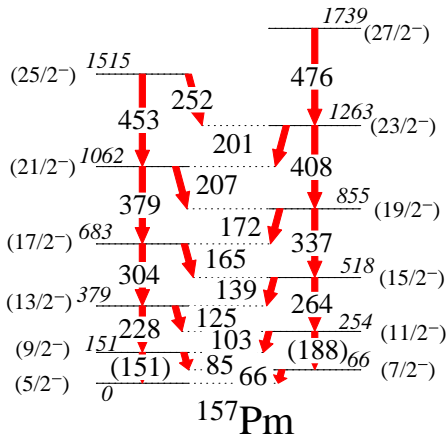}
\caption{\label{fig:scheme-157}  
(Colour online) Level scheme of $^{157}$Pm proposed in the present work.
All the new transitions in the scheme are from the 
present work and are shown in red colour.
The spin-parity of the levels shown are only tentative, assigned  
from the systematics of odd-A Pm isotopes.               }
\end{figure}

The excited states of $^{157}$Pm are identified for the first time from the present 
measurements. 
Some of the transitions were earlier assigned to $^{156}$Pm~\cite{Hwang}. 
The incorrect placement was mainly due to problem of identification 
only from the 
high fold $\gamma$ coincidence data for such exotic nuclei with a small production
 cross section. In the present work, the particular fragment is directly
isotopically  identified using the VAMOS++ spectrometer and the corresponding 
Doppler corrected 
$\gamma$-rays are detected by EXOGAM segmented Clover array, leading
to an unambiguous identification of the relevant transitions.
The $\gamma$-ray singles spectrum in coincidence with detected $^{157}$Pm fragments 
is shown in Fig.~\ref{fig:spec-odd}(d).
The $\gamma$-$\gamma$ coincidences for such neutron-rich nucleus 
were not possible from the measurements of $^{238}$U+$^{9}$Be reaction,  
but were obtained from 
$^{252}$Cf fission data of Gammasphere array. The two representative coincidence 
spectra corresponding to double gates of 39~keV (Pm X-ray) and 264~keV,   
39~keV (Pm X-ray) and 103~keV are shown in Fig.~\ref{fig:tripple-157}. 
The proposed level scheme of $^{157}$Pm, obtained from the present work, 
is shown in Fig.~\ref{fig:scheme-157}. 
Only transitions seen in Fig.~\ref{fig:spec-odd}(d) and in
multiple coincidence gates such as in Fig.~\ref{fig:tripple-157}, 
are placed in Fig.~\ref{fig:scheme-157}.
An indication of the transitions 151~keV and 188~keV could be observed
in (A,Z) gated spectra. These transitions are tentatively placed as 
the E2 crossover transitions corresponding to 
(9/2$^-$) to (5/2$^-$) and (11/2$^-$)to (7/2$^-$) states respectively. 
The level scheme is obtained mainly from the energy level systematics of the 
neighbouring isotopes, energy-sum, intensity balance and coincidence information. 
For certain levels (11/2$^-$ state) the observed $\gamma$ 
ray intensities feeding in and decaying out of the level are imbalanced. 
This can be accounted by considering the proper conversion of lower energy 
dipole $\gamma$ rays according to their mixing with higher order multipoles.   
Spin-parity of the states are only tentatively assigned from the systematics 
of odd-A Pm isotopes of lower masses.

\begin{center}

\begin{longtable}{|c|c|ccc|c|}

\caption{\label{tab:Table1}The energies (E$_\gamma$) and relative intensities (I$_\gamma$) of the 
$\gamma$ rays assigned to different odd-A Pm isotopes along with the spin and parity of the initial 
($J^{\pi}_i$) and the final ($J^{\pi}_f$) states and the energy of the 
initial state (E$_i$). } \\

\hline 
$ E_{\gamma}(keV)^{\footnotemark[1]}$ &$E_i (keV)^{\footnotemark[2]}$   &$ J^{\pi}_i $&$ \rightarrow $&$ J^{\pi}_f $&$ I_{\gamma}^{\footnotemark[3]}(Err.) $ \\
\endfirsthead

\multicolumn{6}{c}%
{Table.~\ref{tab:Table1} continued} \\

\hline
$ E_{\gamma}(keV)^{\footnotemark[1]}$ &$E_i (keV)^{\footnotemark[2]}$   &$ J^{\pi}_i $&$ \rightarrow $&$ J^{\pi}_f $&$ I_{\gamma}^{\footnotemark[3]}(Err.) $ \\

\hline
\endhead

\hline 
\multicolumn{6}{|r|}{{Table.~\ref{tab:Table1} Continued}} \\ \hline
\endfoot
\hline \hline
\endlastfoot

\hline
\bf$^{153}$Pm&&&&&\\
\hline
 66 &   66&$ 7/2^{-}    $&$ \rightarrow $&$  5/2^{-}   $  &24	(6)	     \\
 85 &  151&$ 9/2^{-}    $&$ \rightarrow $&$  7/2^{-}   $  &41	(11)	     \\
 102&  253&$ (11/2^{-}) $&$ \rightarrow $&$  9/2^{-}   $  &74	(12)	     \\
 112&  311&$ (11/2^{+}) $&$ \rightarrow $&$  (9/2^{+}) $  &9	(5)	     \\
 124&  376&$ (13/2^{-}) $&$ \rightarrow $&$ (11/2^{-}) $  &100	(9)	     \\
 133&  199&$ (9/2^{+})  $&$ \rightarrow $&$  7/2^{-}   $  &33	(6)	     \\
 136&  512&$ (15/2^{-}) $&$ \rightarrow $&$ (13/2^{-}) $  &98	(8)	     \\
 151&  151&$ 9/2^{-}    $&$ \rightarrow $&$  5/2^{-}   $  &15	(8)	     \\
 160&  311&$ (11/2^{+}) $&$ \rightarrow $&$  9/2^{-}   $  &36	(8)	     \\
 165&  678&$ (17/2^{-}) $&$ \rightarrow $&$ (15/2^{-}) $  &158	(9)$^{\footnotemark[4]}$ \\
 167&  845&$ (19/2^{-}) $&$ \rightarrow $&$ (17/2^{-}) $  &-$^{\footnotemark[4]}$         \\
 178&  430&$ (13/2^{+}) $&$ \rightarrow $&$ (11/2^{-}) $  &20	(6)	     \\
 187&  253&$ (11/2^{-}) $&$ \rightarrow $&$  7/2^{-}   $  &30	(6)	     \\
 192& 1243&$ (23/2^{-}) $&$ \rightarrow $&$ (21/2^{-}) $  &80	(6)	     \\
 206& 1051&$ (21/2^{-}) $&$ \rightarrow $&$ (19/2^{-}) $  &71	(8)	     \\
 212& 1703&$ (27/2^{-}) $&$ \rightarrow $&$ (25/2^{-}) $  &59	(6)	     \\
 226&  376&$ (13/2^{-}) $&$ \rightarrow $&$  9/2^{-}   $  &48	(11)	     \\
 232&  430&$ (13/2^{+}) $&$ \rightarrow $&$  (9/2^{+}) $  &14	(5)	     \\
 248& 1491&$ (25/2^{-}) $&$ \rightarrow $&$ (23/2^{-}) $  &45	(8)	     \\
 260&  512&$ (15/2^{-}) $&$ \rightarrow $&$ (11/2^{-}) $  &76	(9)	     \\
 286& 1989&$ (29/2^{-}) $&$ \rightarrow $&$ (27/2^{-}) $  &12	(3)	     \\
 301&  678&$ (17/2^{-}) $&$ \rightarrow $&$ (13/2^{-}) $  &92	(8)	     \\
 332&  845&$ (19/2^{-}) $&$ \rightarrow $&$ (15/2^{-}) $  &98	(9)	     \\
 373& 1051&$ (21/2^{-}) $&$ \rightarrow $&$ (17/2^{-}) $  &56	(8)	     \\
 398& 1243&$ (23/2^{-}) $&$ \rightarrow $&$ (19/2^{-}) $  &79	(11)	     \\
 440& 1491&$ (25/2^{-}) $&$ \rightarrow $&$ (21/2^{-}) $  &76	(12)	     \\
 460& 1703&$ (27/2^{-}) $&$ \rightarrow $&$ (23/2^{-}) $  &79	(9)	     \\
 498& 1989&$ (29/2^{-}) $&$ \rightarrow $&$ (25/2^{-}) $  &68	(12)	     \\


\hline
\bf$^{155}$Pm&&&&&\\

\hline

 67  &    67&$  (7/2^{-}) $&$\rightarrow $&$ 5/2^{-}   $  &27   (8)    \\
 87  &   154&$  (9/2^{-}) $&$\rightarrow $&$ (7/2^{-}) $  &54   (14) \\
 105 &   259&$ (11/2^{-}) $&$\rightarrow $&$ (9/2^{-}) $  &73   (5) \\
 127 &   386&$ (13/2^{-}) $&$\rightarrow $&$(11/2^{-}) $  &100  (5) \\
 142 &   528&$ (15/2^{-}) $&$\rightarrow $&$(13/2^{-}) $  &92   (5) \\
 169 &   698&$ (17/2^{-}) $&$\rightarrow $&$(15/2^{-}) $  &81   (5) \\
 174 &   872&$ (19/2^{-}) $&$\rightarrow $&$(17/2^{-}) $  &62   (5) \\
 192 &   259&$ (11/2^{-}) $&$\rightarrow $&$ (7/2^{-}) $  &27   (5) \\
 202 &  1286&$ (23/2^{-}) $&$\rightarrow $&$(21/2^{-}) $  &24   (5) \\
 213 &  1084&$ (21/2^{-}) $&$\rightarrow $&$(19/2^{-}) $  &41   (5) \\
 223 &  1769&$ (27/2^{-}) $&$\rightarrow $&$(25/2^{-}) $  &27   (5) \\
 232 &   387&$ (13/2^{-}) $&$\rightarrow $&$ (9/2^{-}) $  &46   (5) \\
 259 &  1546&$ (25/2^{-}) $&$\rightarrow $&$(23/2^{-}) $  &14   (3) \\
 269 &   528&$ (15/2^{-}) $&$\rightarrow $&$(11/2^{-}) $  &51   (5) \\
 311 & 698   &$ (17/2^{-})$&$ \rightarrow$&$ (13/2^{-}) $& 78   (8) \\
 343 & 872   &$ (19/2^{-})$&$ \rightarrow$&$ (15/2^{-}) $& 70   (8) \\
 387 &1084   &$ (21/2^{-})$&$ \rightarrow$&$ (17/2^{-}) $& 46   (5) \\
 415 &1286   &$ (23/2^{-})$&$ \rightarrow$&$ (19/2^{-}) $& 38   (5) \\
 461 &1546   &$ (25/2^{-})$&$ \rightarrow$&$ (21/2^{-}) $& 38   (5) \\
 482 &1769   &$ (27/2^{-})$&$ \rightarrow$&$ (23/2^{-}) $& 41   (5) \\
\hline
\bf$^{157}$Pm&&&&& \\

\hline

 66   &    66&$  (7/2^{-}) $&$ \rightarrow $&$  (5/2^{-}) $ & 2(1)  \\
 85   &   151&$  (9/2^{-}) $&$ \rightarrow $&$  (7/2^{-}) $ & 5(2)  \\
 103 &   254&$ (11/2^{-}) $&$ \rightarrow $&$  (9/2^{-}) $  & 6(1)  \\
 125 &   379&$ (13/2^{-}) $&$ \rightarrow $&$ (11/2^{-}) $  & 9(1)  \\
 139 &   518&$ (15/2^{-}) $&$ \rightarrow $&$ (13/2^{-}) $  & 6(2)  \\
 151 &   151&$ (9/2^{-})  $&$ \rightarrow $&$  (5/2^{-}) $  & -$^{\footnotemark[5]}$     \\
 165 &   683&$ (17/2^{-}) $&$ \rightarrow $&$ (15/2^{-}) $  & 10(1) \\
 172 &   855&$ (19/2^{-}) $&$ \rightarrow $&$ (17/2^{-}) $  & 6(1)  \\
 188 &   254&$ (11/2^{-}) $&$ \rightarrow $&$  (7/2^{-}) $  & 2(1)  \\
 201 &  1263&$ (23/2^{-}) $&$ \rightarrow $&$ (21/2^{-}) $  & 4(1)  \\
 207 &  1062&$ (21/2^{-}) $&$ \rightarrow $&$ (19/2^{-}) $  & 4(1)  \\
 228 &   379&$ (13/2^{-}) $&$ \rightarrow $&$  (9/2^{-}) $  & 2(1)  \\
 252 &  1515&$ (25/2^{-}) $&$ \rightarrow $&$ (23/2^{-}) $  & 2(1)  \\
 264 &   518&$ (15/2^{-}) $&$ \rightarrow $&$ (11/2^{-}) $  & 7(2)  \\
 304 &   683&$ (17/2^{-}) $&$ \rightarrow $&$ (13/2^{-}) $  & 6(1)  \\
 337 &   855&$ (19/2^{-}) $&$ \rightarrow $&$ (15/2^{-}) $  & 6(1)  \\
 379 &  1062&$ (21/2^{-}) $&$ \rightarrow $&$ (17/2^{-}) $  & 2(1)  \\
 408 &  1263&$ (23/2^{-}) $&$ \rightarrow $&$ (19/2^{-}) $  & 3(1)  \\
 453 &  1515&$ (25/2^{-}) $&$ \rightarrow $&$ (21/2^{-}) $  & 4(2)  \\
 476 &  1739&$ (27/2^{-}) $&$ \rightarrow $&$ (23/2^{-}) $  & 7(4)\\

\footnotetext[1]{ $\gamma$-ray energy uncertainties are typically 
$\pm$ 0.2~keV, $\pm$ 0.5~keV and $\pm$ 1~keV around 200~keV, 500~keV and 1~MeV respectvely.}  
\footnotetext[2]{ The maximum uncertainty of the level energy is upto 0.5\%.} 
\footnotetext[3]{Intensities are normalized to 100 for $^{153,155}$Pm and to 10 for $^{157}$Pm. 
The errors quoted are the fitting errors.}
\footnotetext[4]{Combined intensity of 165 and 167~keV.}
\footnotetext[5]{Weak transition, intensity could not be determined.}

\end{longtable}
\end{center}

\subsection{\bf Odd-odd Pm isotopes}

\vskip 0.3cm
\begin{figure}[ht]
\includegraphics[width=\columnwidth]{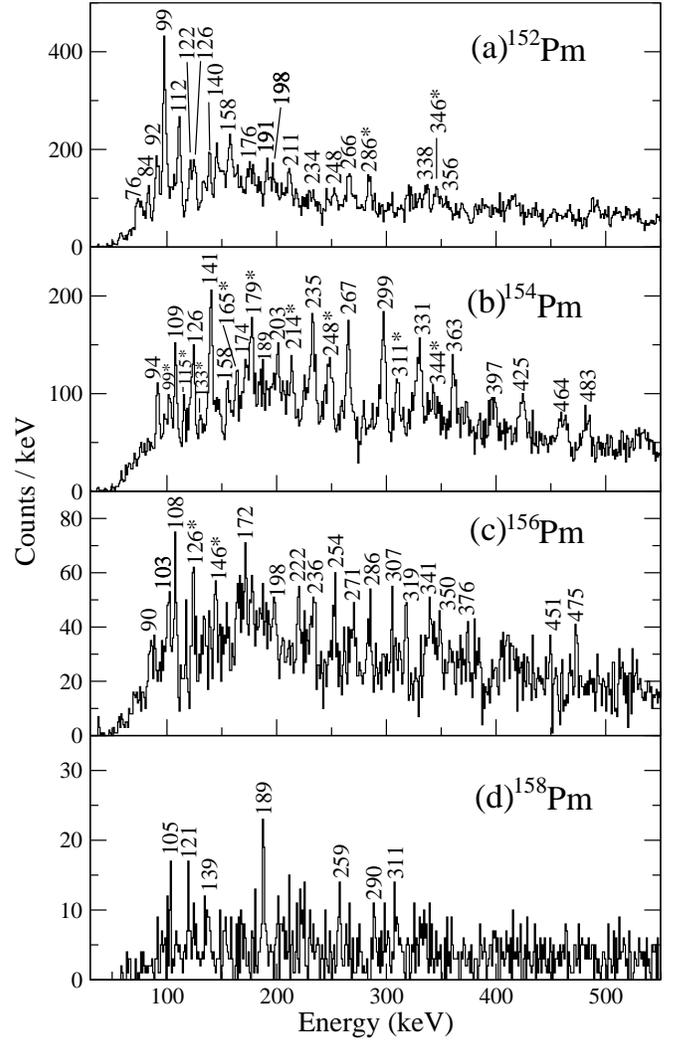}
\caption{\label{fig:spec-even}(A,Z) gated Doppler corrected 
``singles'' $\gamma$ spectra of $^{152-158}$Pm, obtained from  
the fragment-$\gamma$ coincidence in $^{238}$U+$^{9}$Be-induced fission data.
The $\gamma$ rays marked as '*' are assigned to the respective even-A Pm isotopes
from the corresponding (A,Z) coincidence, but could not be 
placed in the level schemes}.
\end{figure}

\begin{figure}[ht]
\includegraphics[width=\columnwidth]{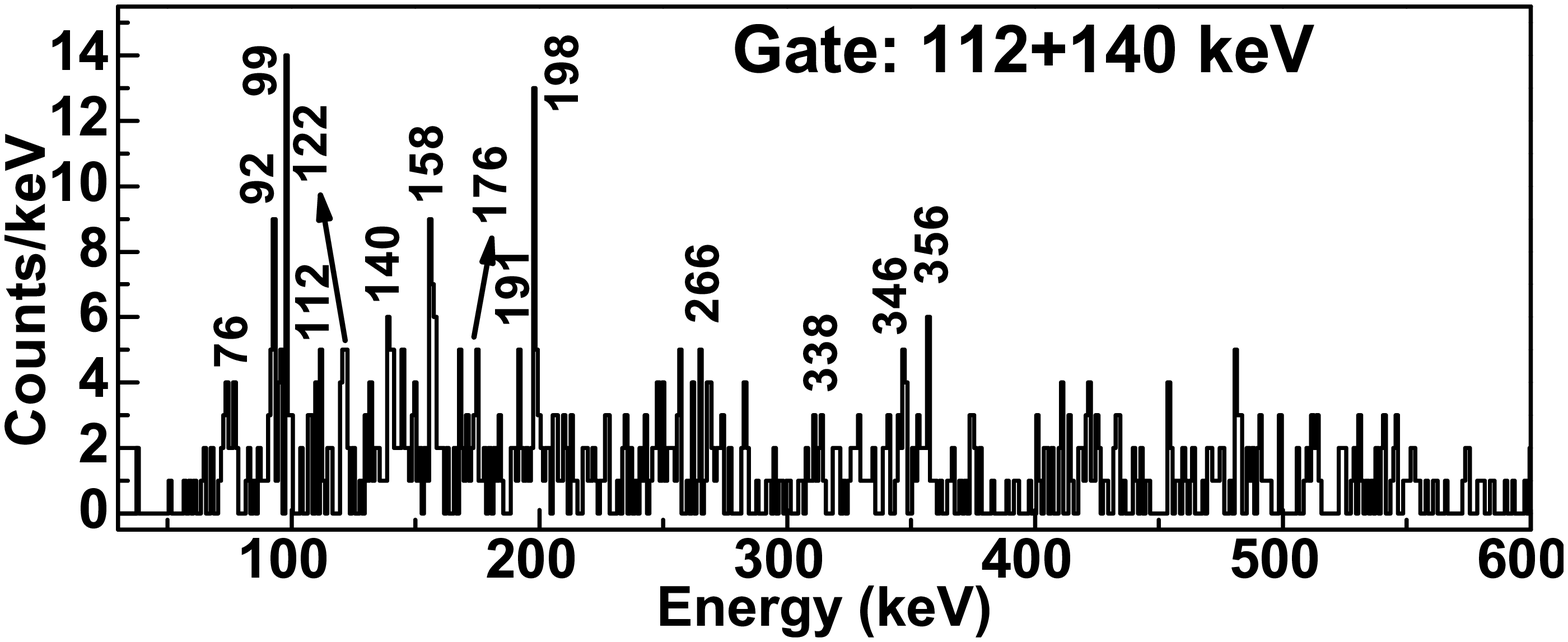}
\caption{\label{fig:152-gg}  
Coincidence spectrum corresponding to the added gates of 112 and 140~keV transitions 
of $^{152}$Pm, obtained from $^{238}$U+$^{9}$Be-induced fission data.  }
\end{figure}  

\begin{figure}[ht]
\includegraphics[width=\columnwidth]{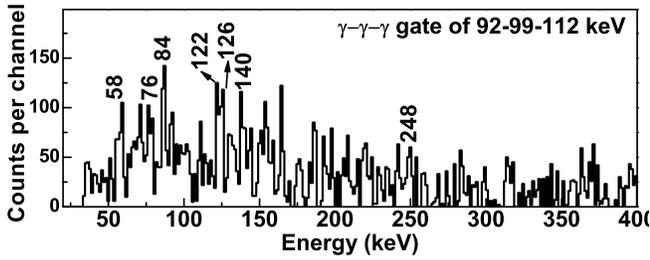}
\caption{\label{fig:152-ggg}  
Coincidence spectrum corresponding to triple gates of 
92, 99 and 112~keV transitions of $^{152}$Pm, 
obtained from $^{252}$Cf fission. }
\end{figure}

\begin{figure}[ht]
\includegraphics[height=0.8\columnwidth, angle=-90]{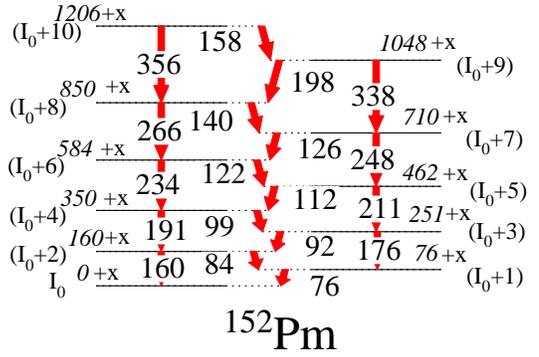}
\caption{\label{fig:scheme-152}  
(Colour online) Level scheme of $^{152}$Pm proposed in the present work.
All the new transitions in the scheme are from the 
present work and are shown in red colour.
The level energies and tentative spin assignments are shown 
with respect to a level of energy (0+x) and spin I$_0$.}
\end{figure}

For odd-odd Pm isotopes, the present work reports the first in-beam measurements 
of prompt $\gamma$ rays of neutron-rich $^{152, 154, 156, 158}$Pm isotopes. 
The Doppler corrected $\gamma$-ray spectra of odd-odd Pm isotopes, 
detected in the present work in coincidence with the respective isotopes identified at 
the focal plane of VAMOS++ spectrometer are shown in Fig.~\ref{fig:spec-even}. 
The $\gamma$ rays which could not be placed in the level schemes of
respective even-A Pm isotopes due to lack of $\gamma$-$\gamma$ 
coincidence information, are labelled in Fig.~\ref{fig:spec-even} 
(see caption of Fig.~\ref{fig:spec-even}).
However, the identification of these transitions to the 
corresponding even-A Pm isotopes are confirmed from (A,Z) gating condition.
All these neutron-rich odd-odd Pm isotopes are known to have 
long-lived isomers and the information on excited states of these isotopes 
were extracted
only from $\beta$-decays. Prior to this study, the information on 
the high spin states of these neutron rich nuclei were not available.
This is because
these states are difficult to populate in reactions other than fission
and unambiguous assignment of the $\gamma$ transitions of such neutron rich nuclei 
cannot be made without direct identification of the corresponding isotope in coincidence.
In particular, the presence of long-lived isomers make the situation more complicated, 
as correlation of the transitions below and above the isomer is not possible only from 
prompt $\gamma$-$\gamma$ coincidence without any isotopic identification.  
 In the present work, only the prompt in-beam $\gamma$-rays are detected 
with isotopic identification as
the experimental setup was not optimized to detect the delayed $\gamma$-rays. 
The $\gamma$ rays, assigned to various even-A Pm isotopes, along with their 
relative intensities, 
obtained from the (A,Z) gated spectra are shown in 
Table.~\ref{tab:Table2}. The probable spin-parity the initial and final states are 
only tentatively assigned from the systematics of even-A Pm isotopes. 
The level energies given in Table.~\ref{tab:Table2} are with respect to a reference level and  
the quoted errors are the fitting errors.
The $\gamma$ rays observed and the proposed level schemes of 
particular cases are discussed below.

The precise measurement of the low spin levels of $^{152}$Pm
were established from the decay of mass separated $^{152}$Nd~\cite{Shibata}. 
Strong $\beta$ decay feeding to the ground state (1$^+$) and other low spin states 
(mostly 1$^+$) of $^{152}$Pm are reported.
The presence of two long-lived isomers with half lives 7.5~min and $\approx$18~min 
were identified in $^{152}$Pm and the most probable spin-parity for these 
isomeric states were proposed to be of 4$^{\pm}$ and $\ge$6$^+$, respectively, 
based on the decay feeding to $^{152}$Sm levels~\cite{William}.
Though the total $\beta$-decay energy of 7.5~min isomer
was found to be 3.6~$\pm$~0.1~MeV, the excitation energy of the proposed 
$\approx$18 min isomer could not be established. 
In the present work 
the prompt $\gamma$ ray spectrum obtained in coincidence with $^{152}$Pm 
fragments is shown in Fig.~\ref{fig:spec-even}(a). 
The information about the coincidence relationships 
of these $\gamma$ rays have been obtained from the limited statistics 
of $\gamma$-$\gamma$ coincidence from $^{238}$U+$^{9}$Be reaction. 
The coincidence spectrum corresponding 
to the sum of gates of 112 and 140~keV $\gamma$ rays is shown in Fig.~\ref{fig:152-gg}. 
The presence of 92-99-112-140-158-198~keV cascade can be seen from this spectrum. 
The (A,Z) gated $\gamma$ rays of $^{152}$Pm (Fig.~\ref{fig:spec-even}(a)) are 
then used to obtain the coincidence information from the high-fold 
$\gamma$-$\gamma$-$\gamma$ 3D cube and $\gamma$-$\gamma$-$\gamma$-$\gamma$ 4D cubes
from the Gammasphere data. 
A representative spectrum corresponding to triple gate of 92-99-112~keV cascade 
is shown in Fig.~\ref{fig:152-ggg}. 
The level scheme, as shown in Fig.~\ref{fig:scheme-152}, is obtained from the 
$\gamma$-$\gamma$ coincidence information and the energy-sum systematics. 
As none of the $\gamma$-rays, reported earlier from the $\beta$-decay of 
$^{152}$Nd~\cite{Shibata} are observed in the present work,  
the excitation energy of the lowest level of $^{152}$Pm from the current work 
cannot be established. 
As mentioned earlier, mainly high spin states are populated in fission reactions. 
As a result, the excited states of $^{152}$Pm populated in the present work most  
probably decay to either of the known high spin isomeric states.
Thus, the excitation energy and spin of the lowest level of the proposed scheme 
as (0+x) and I$_0$ respectively. 

\begin{figure}[ht]
\includegraphics[width=\columnwidth]{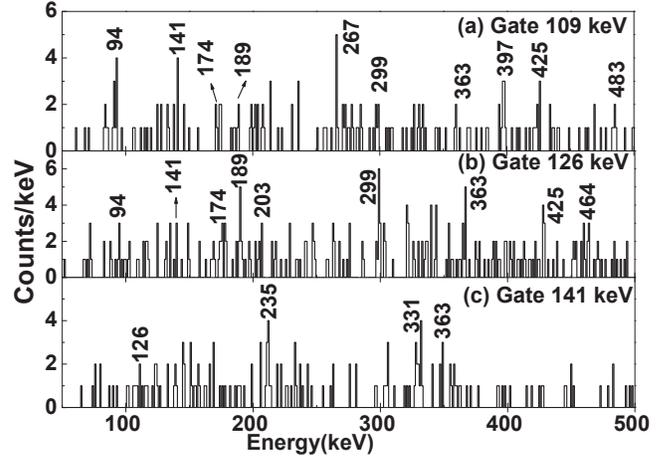}
\caption{\label{fig:154-gg}  
Coincidence spectra corresponding to (a) 109, (b) 126 and (c) 141~keV gates of $^{154}$Pm, 
obtained from $^{238}$U+$^{9}$Be-induced fission data.  }
\end{figure}  

\begin{figure}[ht]
\includegraphics[width=\columnwidth]{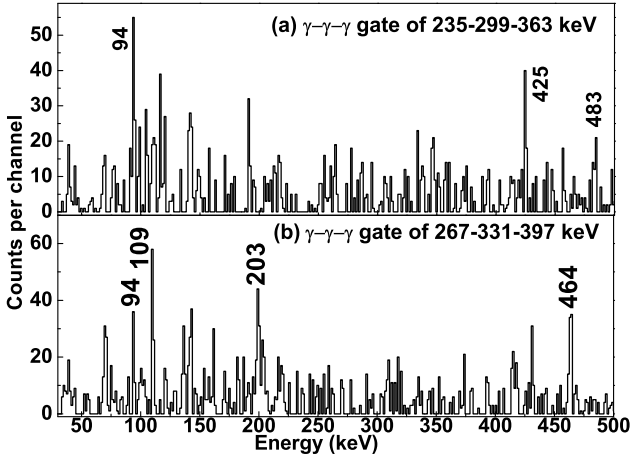}
\caption{\label{fig:154-ggg}  
Coincidence spectra corresponding to triple gates of 
(a) 235-299-363~keV and (b) 267-331-397~keV 
transitions of $^{154}$Pm, 
obtained from $^{252}$Cf fission.  }
\end{figure}

\begin{figure}[ht]
\includegraphics[width=0.6\columnwidth]{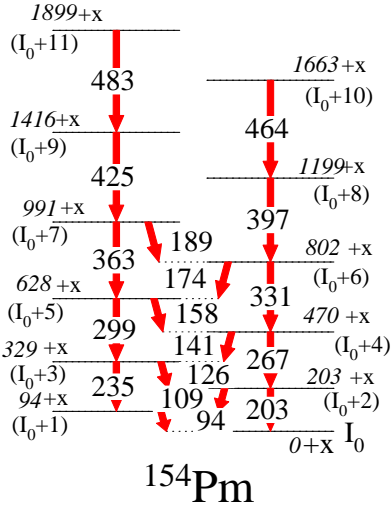}
\caption{\label{fig:scheme-154}  
(Colour online) Level scheme of $^{154}$Pm proposed in the present work.
All the new transitions in the scheme are from the 
present work and are shown in red colour.
The level energies and tentative spin assignments are shown 
with respect to a level of energy (0+x) and spin I$_0$.}
\end{figure}

For $^{154}$Pm, the $\gamma$ ray spectrum in coincidence with $^{154}$Pm 
fragments detected from $^{238}$U+$^{9}$Be reaction, is shown in 
Fig.~\ref{fig:spec-even}(b).
The coincidence spectrum corresponding to some of the gates  
of 109, 126 and 141~keV $\gamma$ rays are shown in Fig.~\ref{fig:154-gg}.
The spectra corresponding to the triple gates obtained from the 
high fold $\gamma$ coincidence data on  spontaneous fission of 
$^{252}$Cf at Gammasphere are shown in Fig.~\ref{fig:154-ggg}. 
The level scheme of $^{154}$Pm proposed in the present work is shown in 
Fig.~\ref{fig:scheme-154}. 
It can be seen from the $\gamma$ spectrum of Fig.~\ref{fig:spec-even}(b) that several 
$\gamma$ rays, such as, 99, 115, 133, 165, 179, 214, 248, 311, 344~keV, 
identified as belonging 
to $^{154}$Pm, are not placed in the proposed level scheme. 
It is possible that these unassigned $\gamma$ rays belong to a different 
band in $^{154}$Pm.
Though the excitation energy of the 2.68 min isomer (3, 4) is not 
known experimentally, this state is considered as the probable ground state from 
the consideration of $\beta$ transition rates and expected two quasi-particle 
configuration ~\cite{Sood1990} compared to the other 
1.73 min isomer of proposed spin (0$^-$, 1$^-$).
Thus, in the case of $^{154}$Pm, the proposed level scheme can most probably 
be built on the 2.68 min (3, 4) state. 
But, as the energy of the 2.68 min (3, 4) state is not established experimentally, 
the lowest level of the proposed scheme of Fig.~\ref{fig:scheme-154} 
is marked as (0+x), I$_0$.

\begin{figure}[ht]
\includegraphics[width=\columnwidth]{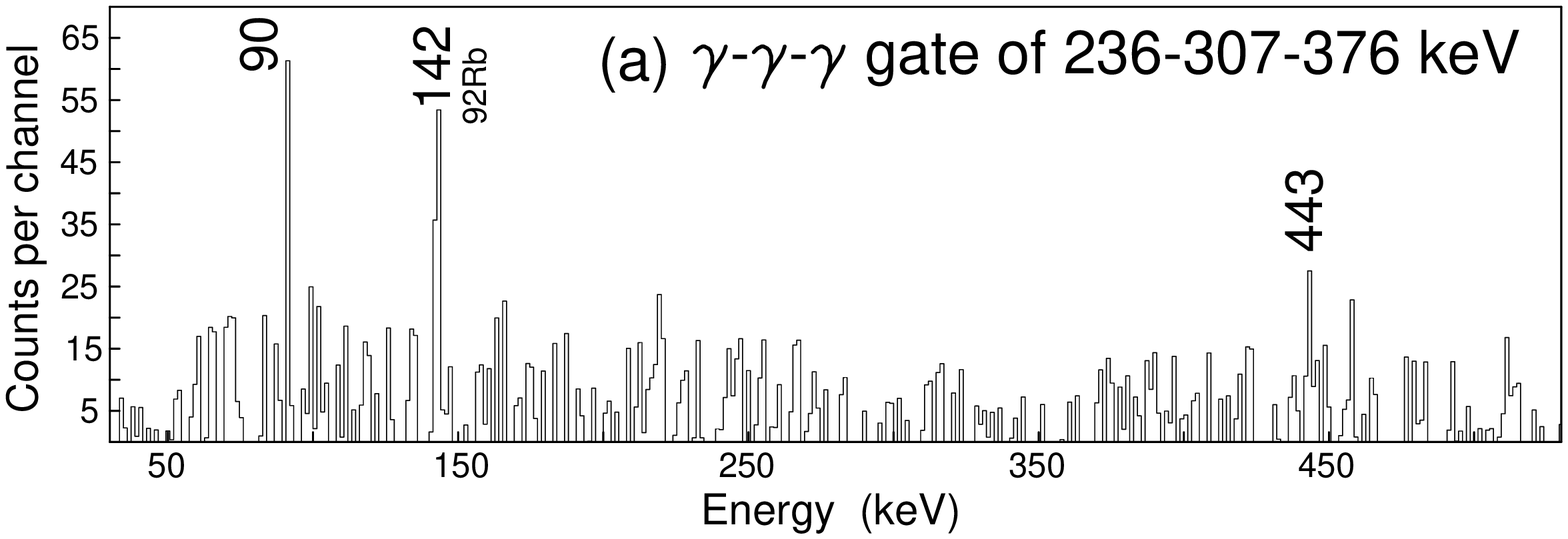}
\includegraphics[width=\columnwidth]{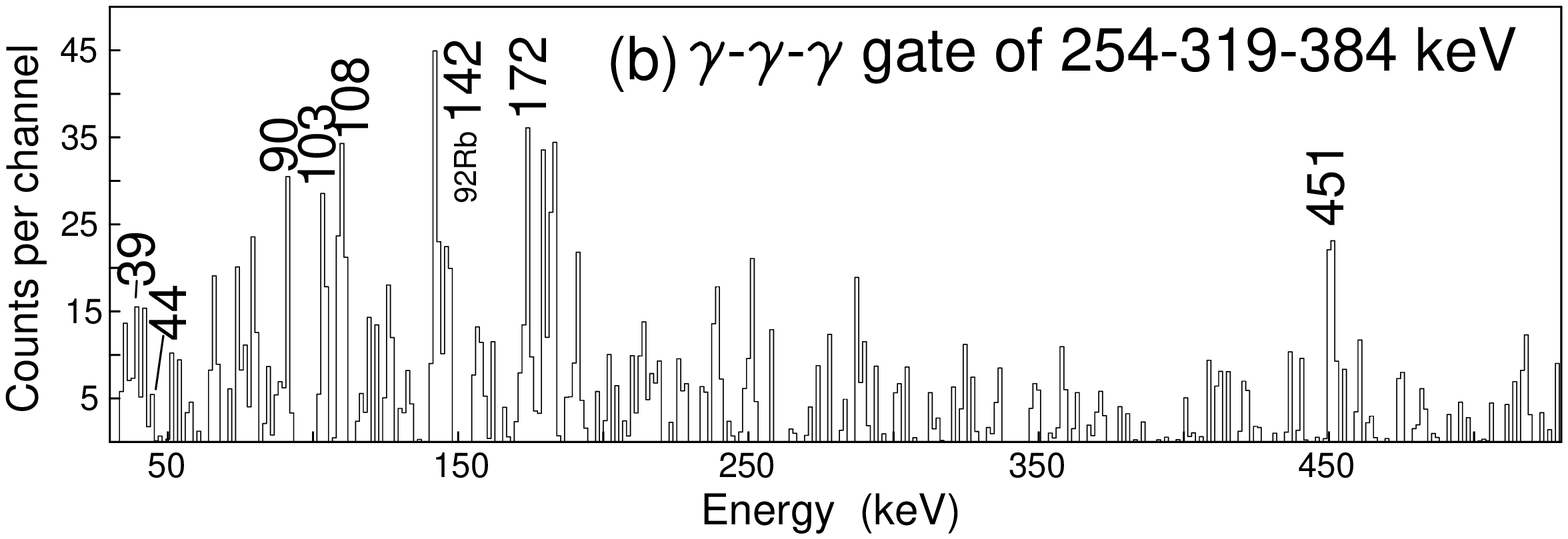}
\includegraphics[width=\columnwidth]{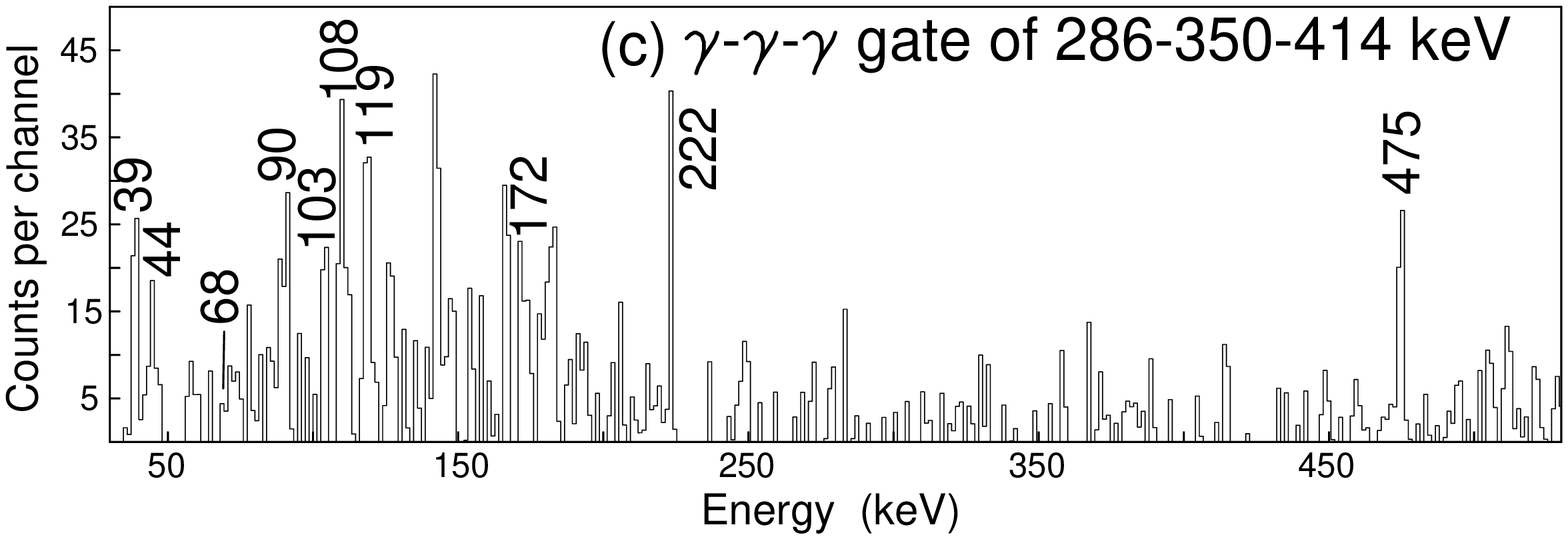}
\caption{\label{fig:156-ggg}  
Coincidence spectra corresponding to triple gates of 
(a) 236-307-376~keV, (b) 254-319-384~keV and 
(c) 286-350-414~keV transitions of $^{156}$Pm, 
obtained from $^{252}$Cf fission.  }
\end{figure}

\begin{figure}[ht]
\includegraphics[height=0.7\columnwidth]{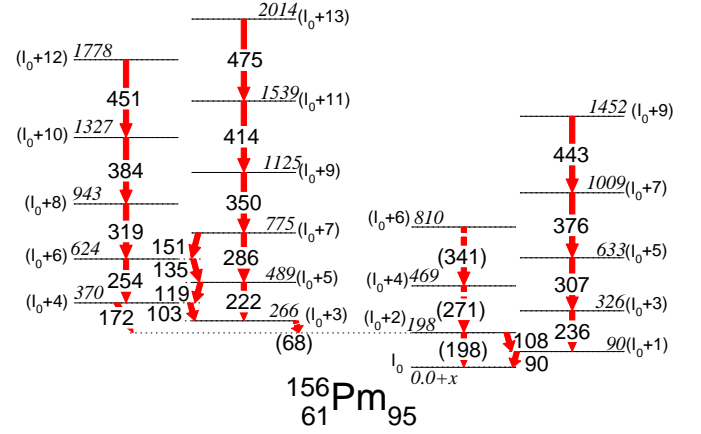}
\caption{\label{fig:scheme-156}  
(Colour online) Level scheme of $^{156}$Pm proposed in the present work. 
All the new transitions in the scheme are from the 
present work and are shown in red colour. 
The level energies and tentative spin assignments are shown 
with respect to a level of energy (0+x) and spin I$_0$. }
\end{figure}

The $\gamma$ ray spectrum of $^{156}$Pm obtained in the present 
work from $^{238}$U+$^{9}$Be reaction, in coincidence 
with $^{156}$Pm fragments identified in the focal plane of VAMOS++ spectrometer
is shown in Fig.~\ref{fig:spec-even}(c). 
It may be noted that none of the $\gamma$ rays assigned to $^{156}$Pm, 
from the previous study of spontaneous fission of $^{252}$Cf by Hwang et 
al.~\cite{Hwang}, could be observed in this spectrum. 
These $\gamma$ rays are now assigned to $^{157}$Pm in the present work,
as described in Section \ref{sec:level02}-A. 
Thus the unique identification of $\gamma$ rays
in case of such an exotic nucleus was difficult from $^{252}$Cf spontaneous fission 
measurements, where the high-fold $\gamma$ coincidences and the cross coincidence 
relationship among the fission fragment partners are utilized to assign the 
$\gamma$ rays to a particular nucleus. 
In the present work, as the fragments are directly identified at the focal 
plane of the spectrometer by (A,Z) tagging, the corresponding coincident 
$\gamma$ rays are uniquely assigned to that particular fragment.
Within the limited statistics of (A,Z) gated data from $^{238}$U+$^{9}$Be reaction, 
no $\gamma$-$\gamma$ coincidence information could be obtained. 
However, the coincidence relationship among various $\gamma$ rays identified  
in (A,Z) gated spectrum could be obtained from the high-fold data from 
the spontaneous fission of $^{252}$Cf, measured using Gammasphere array.
The coincidence spectra corresponding to the triple gates of $\gamma$ transitions 
of $^{156}$Pm are shown in Fig.~\ref{fig:156-ggg}. It may be noted that in these 
coincidence spectra from $^{252}$Cf fission data, the $\gamma$ rays of the 
corresponding fission partner can also be present in coincidence, as there is no
(A,Z) selection in this case. The 142~keV strong transition observed in 
each coincidence spectra of Fig.~\ref{fig:156-ggg}, is actually 
from the fission partner $^{92}$Rb.   
The level scheme, shown in Fig.~\ref{fig:scheme-156}, is obtained from the 
coincidence relationship among various $\gamma$ rays cascades.  
The low energy transition of 68~keV is tentatively placed, as the 
transitions placed above and below the 68~keV are found to be in coincidence.   
Shibata {\it et. al.}~\cite{MShibata} earlier reported $\gamma$ rays 
from the $\beta$-decay of $^{156}$Nd and 
did not propose any level scheme, except for the the isomeric transition decay. 
It is possible that the present level scheme of $^{156}$Pm is built 
on the ground state, which is proposed to be of higher spin (J=4)
compared to the known isomeric state at 150.3~keV.
A ground state of 4$^{(-)}$ was proposed in Ref.~\cite{MShibata}, with the observation of 
a M3 transition from the (1$^-$) isomer. However, in Ref.~\cite{NDS113}, 
the most probable ground state is adopted to be a 4$^{(+)}$ and the isomer as (1$^+$).
From the discussion of Refs.~\cite{MShibata} and ~\cite{NDS113}, it appears that 
configurations corresponding to both 4$^{(+)}$ and 4$^{(-)}$ can be present in this region.
Though some of the $\gamma$ rays reported in Ref.~\cite{MShibata} 
are found to have energies close to those assigned to $^{156}$Pm in the present work, 
it cannot be firmly concluded that the corresponding excited states populated from 
$\beta$ decay of $^{156}$Nd are the same as that observed from the present work. 
Thus the possibility of placing the present level scheme to a state 
above the ground state cannot be ruled out. 
In view of above possibilities, we prefer to keep the lowest state of 
the proposed scheme as (0+x, I$_0$).

The $\gamma$ rays from the excited states of $^{158}$Pm are identified for the 
first time in the present work and the corresponding spectrum 
is shown in Fig.~\ref{fig:spec-even}(d). 
Seven new $\gamma$ rays have been identified as belonging to $^{158}$Pm from data of
$^{238}$U+$^{9}$Be reaction. However, for $^{158}$Pm, no level scheme could be 
obtained, as the statistics is very limited.

\begin{center}
\begin{longtable}{|c|c|ccc|c|}

\caption {\label{tab:Table2} The energies (E$_\gamma$) and relative intensities (I$_\gamma$) 
of the $\gamma$~rays assigned to different even-A Pm isotopes along with the probable spin 
and parity of the initial ($J^{\pi}_i$) and the final ($J^{\pi}_f$) states 
and the energy of the initial state (E$_i$).} \\
\hline

\hline 
$ E_{\gamma}(keV)^{\footnotemark[1]}$ &$E_i (keV)^{\footnotemark[2]}$   &$ J^{\pi}_i $&$ \rightarrow $&$ J^{\pi}_f $&$ I_{\gamma}^{\footnotemark[3]}(Err.) $ \\
\endfirsthead

\multicolumn{6}{c}%
{Table.~\ref{tab:Table2} continued} \\

\hline
$ E_{\gamma}(keV)^{\footnotemark[1]}$ &$E_i (keV)^{\footnotemark[2]}$   &$ J^{\pi}_i $&$ \rightarrow $&$ J^{\pi}_f $&$ I_{\gamma}^{\footnotemark[3]}(Err.) $ \\

\hline
\endhead

\hline 
\multicolumn{6}{|r|}{{Table.~\ref{tab:Table2} Continued}} \\ \hline
\endfoot
\hline 
\hline
\endlastfoot

\hline 

\bf$^{152}$Pm&&&&& \\
\hline
 76 &	 76 &$   (I_0+1) $&$ \rightarrow $&$	(I_0)	$  & 29   (8)    \\
 84 &	 160&$   (I_0+2) $&$ \rightarrow $&$	(I_0+1) $  & 31   (8)	 \\
 92 &	 251&$   (I_0+3) $&$ \rightarrow $&$	(I_0+2) $  & 58   (13)	 \\
 99 &	 350&$   (I_0+4) $&$ \rightarrow $&$	(I_0+3) $  & 100  (15)   \\
 112&	 462&$   (I_0+5) $&$ \rightarrow $&$	(I_0+4) $  & 69   (6)	 \\
 122&	 584&$   (I_0+6) $&$ \rightarrow $&$	(I_0+5) $  & 44   (6)	 \\
 126&	 710&$   (I_0+7) $&$ \rightarrow $&$	(I_0+6) $  & 46   (6)	 \\
 140&	 850&$   (I_0+8) $&$ \rightarrow $&$	(I_0+7) $  & 29   (4)	 \\
 158&	1206&$   (I_0+10)$&$ \rightarrow $&$	(I_0+9) $  & 19   (8)	   \\
 160&	 160&$   (I_0+2) $&$ \rightarrow $&$	(I_0)	$  & 54   (10)	 \\
 176&	 251&$   (I_0+3) $&$ \rightarrow $&$	(I_0+1) $  & 25   (4)	   \\
 191&	 350&$   (I_0+4) $&$ \rightarrow $&$	(I_0+2) $  & 25   (4)	 \\
 198&	1048&$   (I_0+9) $&$ \rightarrow $&$	(I_0+8) $  & 21   (4)	 \\
 211&	 462&$   (I_0+5) $&$ \rightarrow $&$	(I_0+3) $  & 31   (10)   \\
 234&	 584&$   (I_0+6) $&$ \rightarrow $&$	(I_0+4) $  & 21   (6)	 \\
 248&	 710&$   (I_0+7) $&$ \rightarrow $&$	(I_0+5) $  & 23   (6)	 \\
 266&	 850&$   (I_0+8) $&$ \rightarrow $&$	(I_0+6) $  & 40   (6)	 \\
 338&	1048&$   (I_0+9) $&$ \rightarrow $&$	(I_0+7) $  & 38   (10)   \\
 356&	1206&$   (I_0+10)$&$ \rightarrow $&$	(I_0+8) $  & 31   (10)   \\

\hline 

\bf$^{154}$Pm&&&&& \\

\hline
 94 &	  94&$   (I_0+1) $&$ \rightarrow $&$	 (I_0)   $	& 43   (11)	 \\
 109&	 203&$    (I_0+2) $&$ \rightarrow $&$	 (I_0+1) $	& 43   (6)	 \\
 126&	 329&$    (I_0+3) $&$ \rightarrow $&$	 (I_0+2) $	& 46   (9)	 \\
 141&	 470&$    (I_0+4) $&$ \rightarrow $&$	 (I_0+3) $	& 71   (6)	 \\
 158&	 628&$    (I_0+5) $&$ \rightarrow $&$	 (I_0+4) $	& 20   (6)	 \\
 174&	 802&$    (I_0+6) $&$ \rightarrow $&$	 (I_0+5) $	& 29   (11)	 \\
 189&	 991&$    (I_0+7) $&$ \rightarrow $&$	 (I_0+6) $	& 17   (6)	 \\
 203&	 203&$    (I_0+2) $&$ \rightarrow $&$	 (I_0)   $	& 40   (6)	 \\
 235&	 329&$    (I_0+3) $&$ \rightarrow $&$	 (I_0+1) $	& 100  (11)	 \\
 267&	 470&$    (I_0+4) $&$ \rightarrow $&$	 (I_0+2) $	& 83   (9)	 \\
 299&	 628&$    (I_0+5) $&$ \rightarrow $&$	 (I_0+3) $	& 97   (9)	 \\
 331&	 802&$    (I_0+6) $&$ \rightarrow $&$	 (I_0+4) $	& 69   (9)	 \\
 363&	 991&$    (I_0+7) $&$ \rightarrow $&$	 (I_0+5) $	& 63   (9)	 \\
 397&	1199&$    (I_0+8) $&$ \rightarrow $&$	 (I_0+6) $	& 57   (9)	 \\
 425&	1416&$    (I_0+9) $&$ \rightarrow $&$	 (I_0+7) $	& 69   (9)	 \\
 464&	1663&$    (I_0+10) $&$ \rightarrow $&$   (I_0+8) $	& 51   (11)	 \\
 483&	1899&$    (I_0+11) $&$ \rightarrow $&$   (I_0+9) $	& 74   (11)	 \\

\hline 
$^{156}$Pm&&&&& \\

\hline
 90 &	  90  &$   (I_0+1) $&$ \rightarrow $&$     (I_0)   $	& 7  (3)  \\
 103&	 370  &$    (I_0+4) $&$ \rightarrow $&$    (I_0+3) $    & 9  (3) \\
 108&	 198  &$    (I_0+2) $&$ \rightarrow $&$    (I_0+1) $    & 10 (1) \\
 119&	 489  &$    (I_0+5) $&$ \rightarrow $&$    (I_0+4) $	& 6  (1) \\
 135&	 624  &$    (I_0+6) $&$ \rightarrow $&$    (I_0+5) $	& 4  (1) \\
 151&	 775  &$    (I_0+7) $&$ \rightarrow $&$    (I_0+6) $	& -$^{\footnotemark[4]}$ \\
 172&	 370  &$    (I_0+4) $&$ \rightarrow $&$    (I_0+2) $	& 10 (1) \\
 198&	 198  &$    (I_0+2) $&$ \rightarrow $&$    (I_0)   $	& 6  (1) \\
 222&	 489  &$    (I_0+5) $&$ \rightarrow $&$    (I_0+3) $	& 7  (1) \\
 236&	 326  &$    (I_0+3) $&$ \rightarrow $&$    (I_0+1) $	& 6  (1) \\
 254&	 624  &$    (I_0+6) $&$ \rightarrow $&$    (I_0+4) $	& 9  (1) \\
 271&	 469  &$    (I_0+4) $&$ \rightarrow $&$    (I_0+2) $	& 9  (3) \\
 286&	 775  &$    (I_0+7) $&$ \rightarrow $&$    (I_0+5) $	& 9  (1) \\
 307&	 633  &$    (I_0+5) $&$ \rightarrow $&$    (I_0+3) $	& 7  (1) \\
 319&	 943  &$    (I_0+8) $&$ \rightarrow $&$    (I_0+6) $	& 10 (1) \\
 341&	 810  &$    (I_0+6) $&$ \rightarrow $&$    (I_0+4) $	& 6  (1) \\
 350&	1125  &$    (I_0+9) $&$ \rightarrow $&$    (I_0+7) $	& 10 (4) \\
 376&	1009  &$    (I_0+7) $&$ \rightarrow $&$    (I_0+5) $	& 6  (1) \\
 384&	1327  &$    (I_0+10) $&$ \rightarrow $&$   (I_0+8) $	& 7  (1) \\
 414&	1539  &$    (I_0+11) $&$ \rightarrow $&$   (I_0+9) $	& 10 (4) \\
 443&	1452  &$    (I_0+9) $&$ \rightarrow $&$    (I_0+7) $	& 3  (1)   \\
 451&   1778  &$    (I_0+12) $&$ \rightarrow $&$   (I_0+10)$    & 6  (1)  \\
 475&   2014  &$    (I_0+13) $&$ \rightarrow $&$   (I_0+11)$    & 9  (1) \\

\footnotetext[1]{ $\gamma$-ray energy uncertainties are typically 
$\pm$ 0.2~keV, $\pm$ 0.5~keV and $\pm$ 1~keV around 200~keV, 500~keV and 1~MeV respectvely.}  
\footnotetext[2]{ Level energies are given with respect to a level of energy x~keV. 
The maximum uncertainty of the level energy is upto 0.5\%.} 
\footnotetext[3]{Intensities are normalized to 100 for $^{152,154}$Pm and to 10 for $^{156}$Pm. 
The errors quoted are the fitting errors.}
\footnotetext[4]{Weak transition, intensity could not be determined.}

\end{longtable}
\end{center}

\section{\bf Discussion}
\label{sec:level03}

The nuclei under study are located at the boundary of octupole deformed
lanthanide region (see Refs.\ \cite{Butler,Agb16}). If located in the
vicinity of the respective Fermi levels, the deformed orbitals emerging from 
the proton $d_{5/2}$ and $h_{11/2}$ spherical subshells and from neutron 
$f_{7/2}$ and $i_{13/2}$ spherical subshells are responsible for possible 
occurrence of reflection asymmetric shapes. These spherical subshells
satisfy the $\Delta j = \Delta l =3$ condition which leads to an increase
of octupole correlations. The predictions of model calculations for the 
position of this boundary depend on the underlying mean field and its 
parametrization (see discussion in Sec.\ IV of Ref.\ \cite{Agb16}). For example, 
the highest neutron number for which covariant density functional calculations 
predict octupole deformation in the ground states of even-even Nd ($Z=60$) 
nuclei is $N=90$ for most of covariant energy density functionals. On the 
contrary, microscopic+macroscopic calculations and Hartree-Fock calculations 
with finite range Gogny D1S force place this number at $N=88$. Octupole 
deformation is even less pronounced in even-even Sm ($Z=62$) nuclei; 
most model calculations place the boundary of the region of octupole 
deformation at $N=88$ \cite{Agb16}. Note that even these nuclei are 
very soft in octupole deformation with very little gain in binding due
to octupole deformation. Thus they are transitional in nature and octupole
dynamical correlations (vibrations) are expected to play an important
role in their structure.

\begin{figure}[t]
\includegraphics[width=\columnwidth]{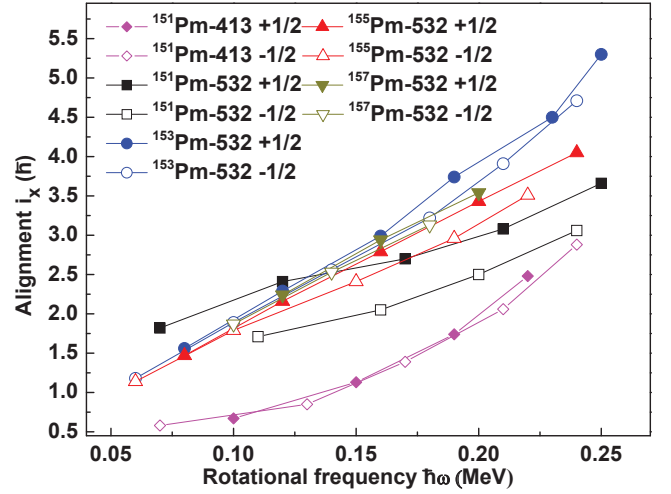}
\caption{\label{fig:ix_Pm}  
(Colour online) Alignment plot for bands of all odd-A Pm isotopes from $N=90$ to $N=96$, 
produced in the present work. 
The different bands are marked by possible Nilsson configurations and 
corresponding signatures ($\alpha$). }
\end{figure}   

  However, the situation becomes more complicated in odd and odd-odd 
nuclei. There are two factors which can stabilize the octupole 
deformation in odd and odd-odd nucleus even if its even-even core 
does not have static octupole deformation. These are polarization 
effects of unpaired particles in specific Nilsson orbitals \cite{Afana95} 
and rotation \cite{Naz92}. Indeed, 
the 5/2[413] and 5/2[523] orbitals, located in the vicinity of the proton
Fermi level of the Pm isotopes of interest, couple strongly
through the $Y_{30}$ operator \cite{Naz92}. However, polarization
energies towards octupole deformation are weak for these
orbitals for neutron numbers $N\geq 90$. So, neutron rich
Pm isotopes are not expected to show static octupole deformation
at low spin. On the other hand, the static octupole deformation
can be stabilized by rotation even if the nucleus is only
octupole soft at spin zero \cite{Naz92}. Experimental data
on parity doublet bands in $^{151}$Pm and $^{153}$Eu formed 
by the 5/2[523] and 5/2[413] rotational structures show
all features of the approach of static octupole deformation
at highest observed spins \cite{Ana1993}.

\begin{figure}[t]
\includegraphics[width=\columnwidth]{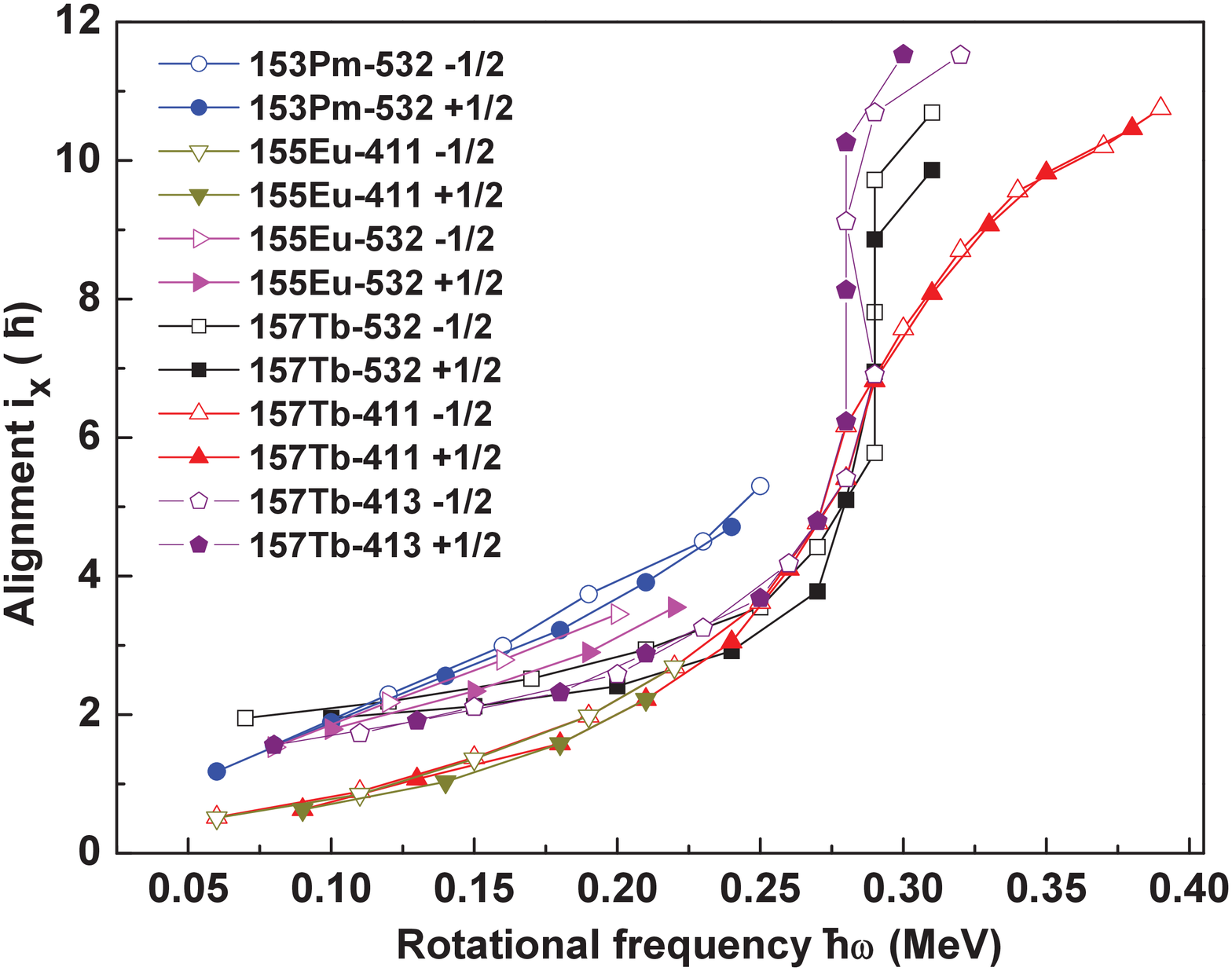}
\caption{\label{fig:ix_Pm_Tb}  
(Colour online) Alignment plot for bands of all odd-A Pm isotopes from $N=90$ to $N=96$, 
produced in the present work. 
The possible configurations of the bands are as mentioned.  }
\end{figure}

\begin{figure}[t]
\includegraphics[width=\columnwidth]{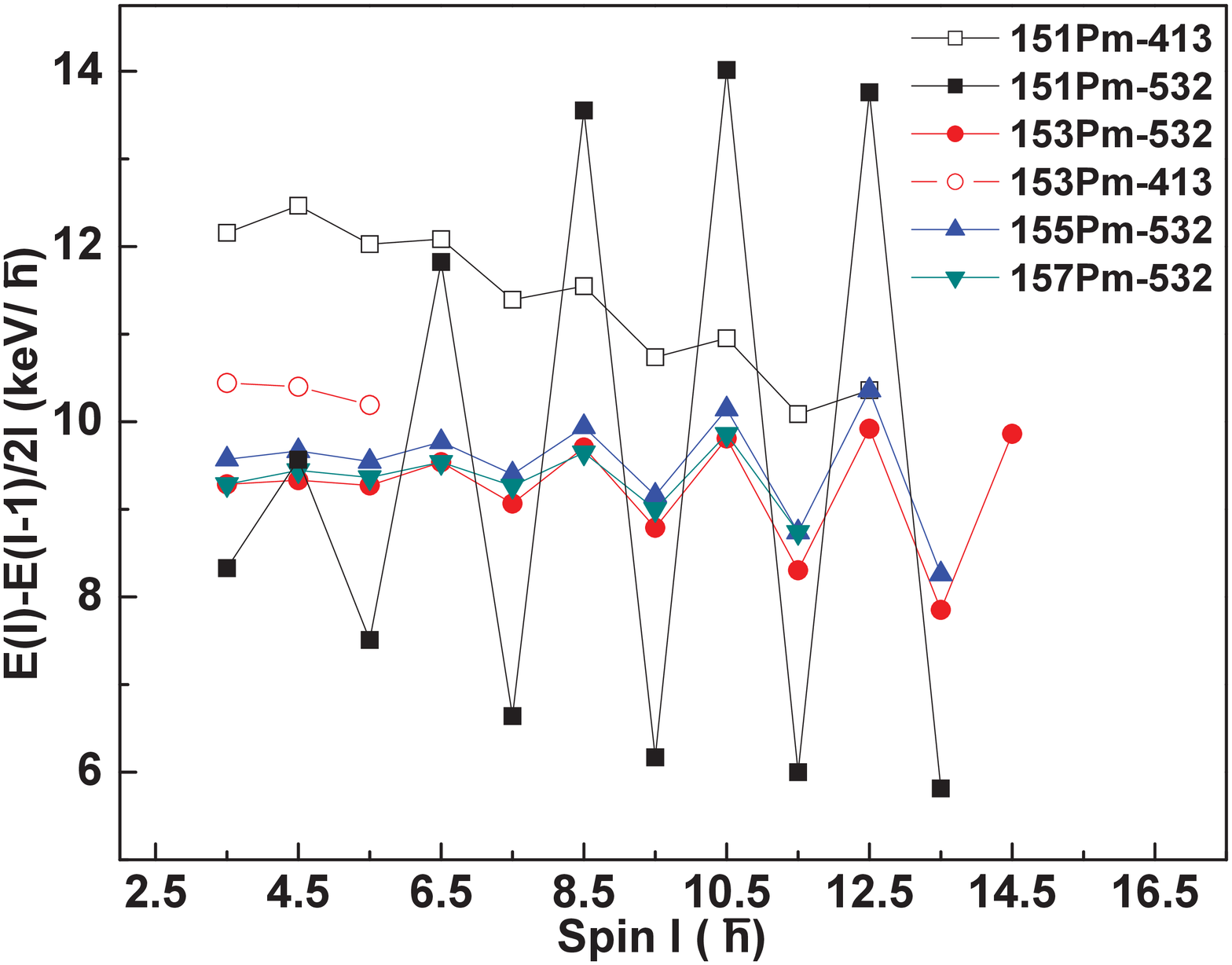}
\caption{\label{fig:stag_Pm}  
(Colour online) Energy staggering as a function of spin for the bands of all odd-A Pm 
isotopes from $N=90$ to $N=96$, produced in the present work. }
\end{figure} 

\begin{figure}[ht]
\includegraphics[width=\columnwidth]{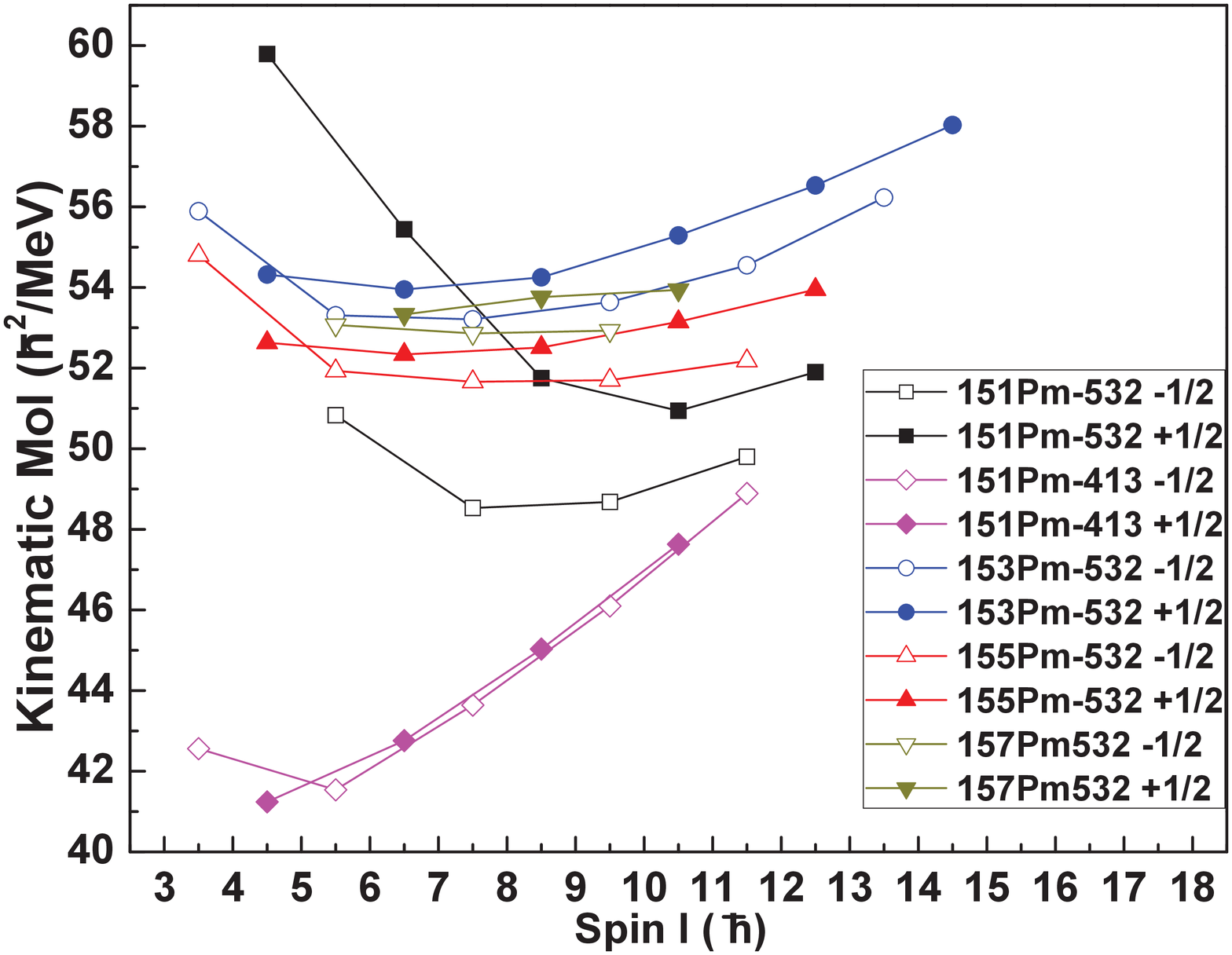}
\caption{\label{fig:KMI_Pm}  
(Colour online) Kinematic moment of inertia (MoI) as a function of spin of the bands of all 
odd-A Pm isotopes from $N=90$ to $N=96$, produced in the present work. }
\end{figure}

In $^{151}$Pm, the opposite parity states of these two rotational sequences
are connected by E1 transitions, which was interpreted as due to the 
presence of static octupole deformation~\cite{Urban}. 
The ground state band in $^{151}$Pm is a positive-parity K= 5/2$^+$  
rotational band based on a 5/2[413] Nilsson orbital originating from $\pi$g$_{7/2}$. 
The negative parity band  built on a K= 5/2$^-$ at an excitation energy of 117~keV is 
based on a 5/2$^-$[532] Nilsson configuration originating from $\pi$$h_{11/2}$ orbital.

\begin{figure*}[t]
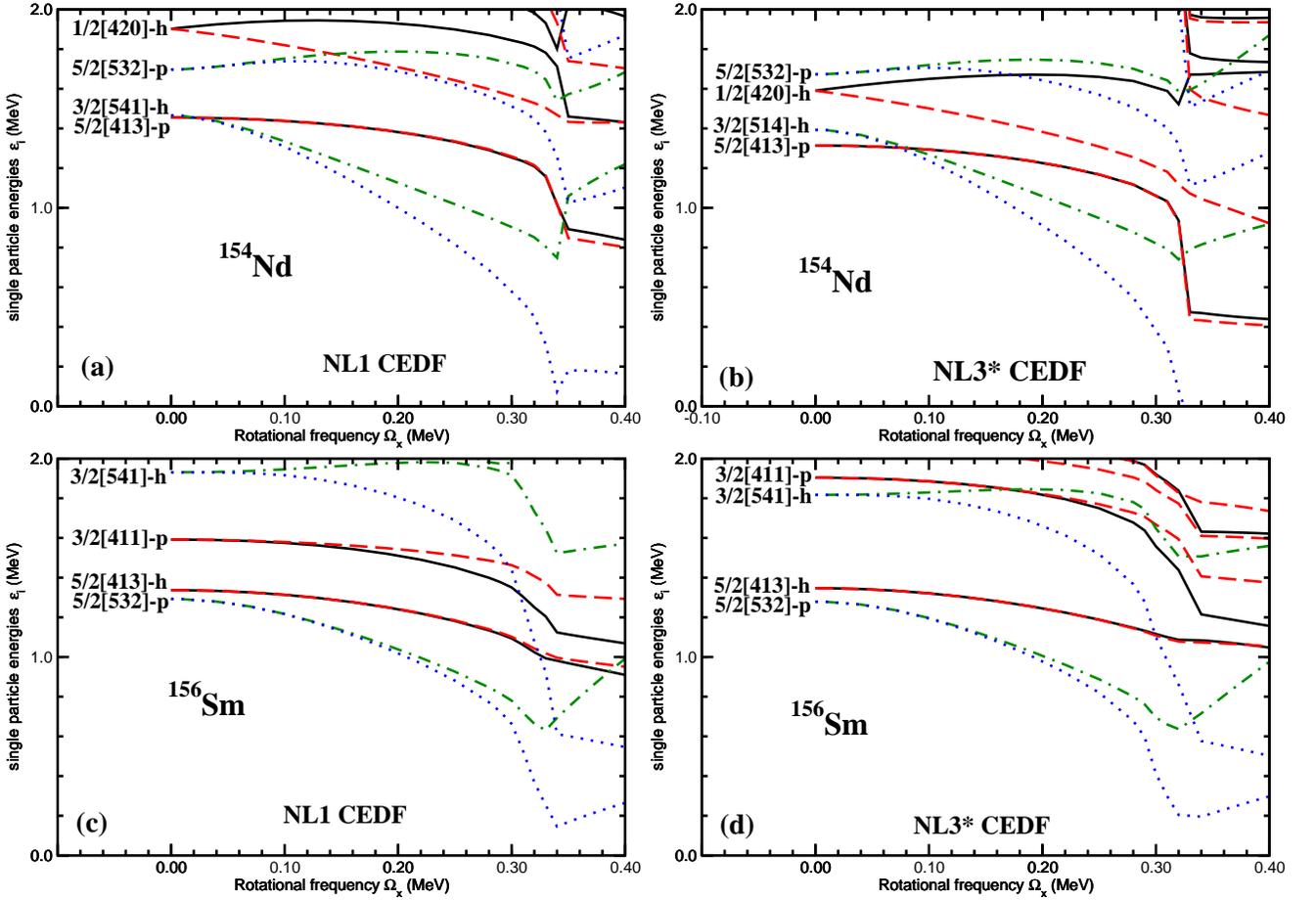

\includegraphics[width=\columnwidth]{Fig22a}
\includegraphics[width=\columnwidth]{Fig22b}
\includegraphics[width=\columnwidth]{Fig22c}
\includegraphics[width=\columnwidth]{Fig22d}
\caption{\label{fig:routhians} (Colour online) Proton quasiparticle energies corresponding 
to the lowest configurations in the indicated nuclei. They are given along 
the deformation path of these configurations. The CRHB+LN calculations 
have been performed  with the NL1 (left panels) and NL3* (right panels) CEDFs. 
Long-dashed, solid, dot-dashed and dotted lines indicate  $(\pi=+, r=+i)$, 
$(\pi=+, r=-i)$, $(\pi=-, r=+i)$ and $(\pi=-, r=-i)$ orbitals, respectively.  
At $\Omega_x=0.0$ MeV, the quasiparticle orbitals are labeled by the asymptotic 
quantum numbers $[Nn_z\Lambda]\Omega$ (Nilsson quantum numbers) of the 
dominant component of the wave function. The letters 'p' and 'h' before the 
Nilsson labels are used to indicate whether a given Routhian is of particle
($V^2 <0.5$) 
or hole ($V^2 > 0.5$) type.}
\end{figure*}

In case of $^{153}$Pm, the yrast ground state band built on K= 5/2$^-$ is known to be 
constituted from 5/2$^-$[532] Nilsson orbital~\cite{Burke}, which originates from
the deformation driving $\pi$$h_{11/2}$ orbital.
In case of $^{151}$Pm, this is the non-yrast structure and the yrast ground band 
is 5/2$^+$ based on 5/2$^+$[413] configuration.
Indeed, in the framework of rotation-vibration model, the level structure of 
$^{153}$Pm, populated by beta decay, is described as that of a deformed nucleus
~\cite{Taniguchi}.
The single proton hole strength in $^{153}$Pm was found to be largely fragmented 
from the study of transfer reactions~\cite{Lee}. 
In the present work, the excited band based on a 5/2$^+$[413] configuration
in $^{153}$Pm could not be extended to higher spin and is weakly populated. 
It is probably due to the fact that the deformation (quadrupole) driving effect 
of the $\pi$$h_{11/2}$ orbital makes $^{153}$Pm to have larger prolate deformation 
compared to $^{151}$Pm.

To understand the band structure in all odd-A Pm isotopes produced in the present work, 
the alignments~\cite{Bengtsson} of both favoured and unfavored signature partner bands are 
plotted in Fig.~\ref{fig:ix_Pm} as a function of rotational frequency. 
The Harris parameters~\cite{Haris} with 
{\it J$_0$}=34.3 ${\hbar}^2$/MeV and {\it J$_1$}=45.0 ${\hbar}^4$/MeV$^3$ have been used 
to subtract the contributions of core angular momentum, which is considered as 
$^{156}$Gd in this case. From the alignment plot of Fig.~\ref{fig:ix_Pm}, 
it is evident that the slopes of the alignments for the bands in 
$^{151}$Pm ($N=90$) are significantly different compared to all  
other odd-A Pm isotopes with higher N/Z. This is due to the fact that 
the nature of the bands in $^{151}$Pm is different as compared to corresponding bands 
in odd-A Pm isotopes with higher $N$. The higher alignment at higher frequency 
in case of odd-A Pm 
isotopes $^{153-157}$Pm compared to $^{151}$Pm is evident as due to the involvement of high-j 
$h_{11/2}$ orbital. The alignments of the bands in $^{155-157}$Pm are quite 
similar to that of the ground band in $^{153}$Pm, which is based on the 5/2[532] configuration 
originating from the $h_{11/2}$ orbital. 
This suggests the same 5/2[532] configuration assignment 
to the observed negative parity bands in $^{155,157}$Pm.
In Fig.~\ref{fig:ix_Pm_Tb} the alignments of the K=5/2$^+$ and K=5/2$^-$ bands in 
$^{153}$Pm are compared with the bands having same configurations and 
with other configurations in neighbouring $N=92$ isotones of $^{157}$Tb and $^{155}$Eu.
In case of $^{157}$Tb, the 3/2$^+$[411] band was found to be the ground state 
band~\cite{Hartley}, whereas in $^{155}$Eu 5/2$^+$[413] band becomes the ground state 
band. The 5/2$^-$[532] band, which is the ground band in $^{153}$Pm was also observed in 
both $^{155}$Eu and $^{157}$Tb. For the 5/2[532] band in $^{157}$Tb, the backbending 
occurs at the frequency $\hbar\omega \sim 0.28$ MeV due to the alignment of two
$i_{13/2}$ neutrons. It can be seen that the same band in $^{153}$Pm could 
be observed in the present work up to the frequency corresponding to the 
upbend of the alignment. 

The energy staggerings of the states in rotational bands of odd-$A$ $^{151-157}$
isotopes are shown in Fig.~\ref{fig:stag_Pm} as a function of spin. From the observed 
signature splitting it is evident that the negative parity band in $^{151}$Pm, 
corresponding to the 5/2$^-$[532] orbital shows pronounced splitting compared to the 
positive parity band corresponding to 5/2$^+$[413] Nilsson orbital. The odd-A Pm 
isotopes with higher N/Z show modest signature splitting 
at higher spins. 
The kinematic moment of inertia ({\it J$_1$}) of the bands in odd-A Pm isotopes 
for both favoured and unfavoured signature partners are plotted in  
Fig.~\ref{fig:KMI_Pm}. The moments of inertia of 5/2$^-$[532] bands in $^{153,155,157}$Pm
isotopes increase at higher spins. 

In order to better understand the experimental features of the observed bands,
proton quasiparticle routhian diagrams for even-even nuclei neighboring to 
$^{155}$Pm are shown in Fig.\ \ref{fig:routhians}. They have been obtained in
the cranked Relativistic Hartree-Bogoliubov calculations (Ref.\ \cite{CRHB}) 
employing two covariant energy density functionals (CEDFs), namely, NL1 \cite{NL1} and NL3* 
\cite{NL3*}. Although the calculated quasiparticle energies somewhat depend on the
employed functional, the rotational features of the routhians of interest
are independent on its choice. For example, in all panels of Fig.\ 
\ref{fig:routhians} the lowest positive parity routhians are based on the
5/2[413] orbital. These routhians are signature degenerated which is similar
to the properties of the experimental 5/2[413] bands seen in $^{151,153}$Pm nuclei. 
The lowest two calculated pairs of negative parity routhians are based on 
the 5/2[532] and 3/2[514] orbitals and their rotational properties depend
on the position of the proton Fermi surface. This surface is more bound in
the $Z=60$ Nd nuclei. As a consequence, the 3/2[514] routhians are the lowest
in energy. The interaction of hole type 3/2[514] and particle type 5/2[532]
orbitals leads to substantial signature splitting in both orbitals. However,
with increasing the energy of the Fermi surface on going to the $Z=62$ Sm isotopes
the 5/2[532] orbital becomes the lowest in energy negative parity orbital and the
coupling between the above mentioned negative parity orbitals is significantly
reduced. As a consequence, the 5/2[532] orbital is signature degenerated at
very low frequencies but small signature degeneracy gradually develops with
increasing rotational frequency. This feature is very similar to what is seen 
in experimental 5/2[532] bands of $^{153,155,157}$Pm (see Figs.\ \ref{fig:ix_Pm_Tb},  
\ref{fig:stag_Pm} and \ref{fig:KMI_Pm}). In addition to the possible role of 
static octupole 
deformation in the structure of the 5/2[532] and 5/2[413] bands of $^{151}$Pm 
discussed in Ref.\ \cite{Ana1993}, it is quite likely that large signature splitting 
seen in the 5/2[532] band of this nuclei is the consequence of the interaction 
of the  5/2[532] and 5/2[413] orbitals (seen in the upper panels of Fig.\ 
\ref{fig:routhians}) which has been discussed above.

 For the odd-odd Pm isotopes, the low lying high-spin isomers are interpreted  
in the framework of quasiparticle-rotor model as two-quasiparticle structures
~\cite{Sood1990, Sood2011}.
The proposed band structures in the present work can either be built just above 
the isomeric state, or above another excited state, which may exist very close 
to the isomeric level. Due to the lack of firm spin and parity assignment
to the band structure built above the long lived isomers, reasonable 
configuration assignments to the observed bands cannot be made.  

\section{\bf Summary and Conclusions}
\label{Concl}

In summary the neutron-rich $^{152-158}$Pm isotopes have been
characterized using in-beam prompt $\gamma$-ray spectroscopy of isotopically
identified fission fragments and high fold coincidence 
data of $^{252}$Cf spontaneous fission. New results of odd-odd Pm isotopes above 
the long lived isomeric states have been reported for the first time.
The rotational band structures of odd-A Pm isotopes with neutron numbers
up to $N=96$ have been extended to higher spins. The 
configuration asignment to the 
rotational structures in odd-A isotopes are understood from the 
systematics of band properties and in terms of routhians of the 
neighbouring even-even isotopes, obtained from 
cranked Relativistic Hartree-Bogoliubov calculations. 
The observed band structures of odd-A Pm isotopes do not show 
any indication of presence of octupole deformation beyond $N=90$.

\section{\bf Acknowledgments}

We would like to thank J.~Goupil, G.~Fremont, L.~M\'{e}nager, J.~Ropert, C.~Spitaels,
and the GANIL accelerator staff for their technical contributions and C.~Schmitt 
for help in various aspects of data collection, analysis and many useful discussions. 
The authors would also like to thank the referee for a very critical reading 
of the manuscript and for useful suggestions in improving the clarity of the manuscript.
Two of us (S.B and S.B) acknowledges partial  financial support 
through the LIA France-India agreement. 
The work at Vanderbilt University and Lawrence Berkeley
National Laboratory are supported by the U.S. Department of
Energy under Grant No. DE-FG05-88ER40407 and Contract
No. DE-AC03-76SF00098. The work at Tsinghua University
was supported by the National Natural Science Foundation
of China under Grant No. 11175095.
The work at JINR was partially supported by the 
Russian Foundation for Basic Research Grant No.
08-02-00089 and by the INTAS Grant No. 03-51-4496.
This material is based upon work supported 
by the U.S. Department of Energy, Office of Science, Office of Nuclear Physics under 
Award No. DE-SC0013037 (Mississippi State University).

\end{document}